\documentclass[pra,twocolumn,floatfix,showpacs,preprintnumbers,groupeaddress,amsmath,amssymb]{revtex4-1}
\usepackage[dvips]{graphics}

\usepackage{calrsfs}
\DeclareMathAlphabet{\pazocal}{OMS}{zplm}{m}{n}
\usepackage[utf8]{inputenc}
\usepackage{mathptmx}
\usepackage{psfrag}
\usepackage{graphicx}						
\graphicspath{{figures/}}
\usepackage[english]{babel}
\newcommand{\bra}[1]{\ensuremath{\left\langle#1\right|}}
\newcommand{\ket}[1]{\ensuremath{\left|#1\right\rangle}}

\usepackage{color}
\definecolor{dred}{rgb}{.6,.0,0.}
\definecolor{dblue}{rgb}{.0,.0,0.6}

\usepackage{tikz}
\usetikzlibrary{shapes.geometric}% für ellipse

\colorlet{mfarbe}{magenta}
\usepackage{bbm}

\begin{document}

\title{Time-Correlated Blip Dynamics of Open Quantum Systems}

\author{Michael Wiedmann}
\author{Jürgen T. Stockburger}
\author{Joachim Ankerhold}
\affiliation{Institute for Complex Quantum Systems and IQST, University of Ulm, 89069 Ulm, Germany}

\date{\today}

\begin{abstract}
 The non-Markovian dynamics of open quantum systems is still a challenging task, particularly in the non-perturbative regime at low temperatures. While the Stochastic Liouville-von Neumann equation (SLN) provides a formally exact tool to tackle this problem for both discrete and continuous degrees of freedom, its performance deteriorates for long times due to an inherently non-unitary propagator. Here we present a scheme which combines the SLN with projector operator techniques based on finite dephasing times, gaining substantial improvements in terms of memory storage and statistics. The approach allows for systematic convergence and is applicable in regions of parameter space where perturbative methods fail, up to the long time domain.  Findings are applied to the coherent and incoherent quantum dynamics of two- and three-level systems. In the long time domain sequential and super-exchange transfer rates are extracted and compared to perturbative predictions.
\end{abstract}


% 03.65.Yz Decoherence; open systems; quantum statistical methods
% 05.40.-a Fluctuation phenomena, random processes, noise, and Brownian motion
% 82.20.Xr Quantum effects in rate constants (tunneling, resonances, etc.)

\maketitle

\section{Introduction}
General theories of open quantum dynamics as introduced in
\cite{breue02, weiss12} provide the mathematical pathway to the
characterization of real-world quantum mechanical systems, subject to
dissipation and dephasing by environmental interactions. Such effects
are crucial across a multitude of fields ranging from solid-state to
chemical physics, quantum optics, and mesoscopic physics.

In the context of a classical environment, linear dissipation can be
modeled by the Caldeira-Leggett oscillator approach \cite{calde8183},
closely linked to Langevin equations \cite{zwan01} which offer a
concise formalism based on retarded friction kernels and Gaussian
random forces (thermal noise). The quantum analogue of friction,
however, needs a much more subtle treatment since it typically creates
substantial correlations between system and environment. Moreover, quantum
fluctuations of a thermal reservoir are non-zero at any temperature,
leading to interesting phenomena and non-trivial ground states.

\begin{figure}[h]
\begin{center}
 \includegraphics[width = 9 cm]{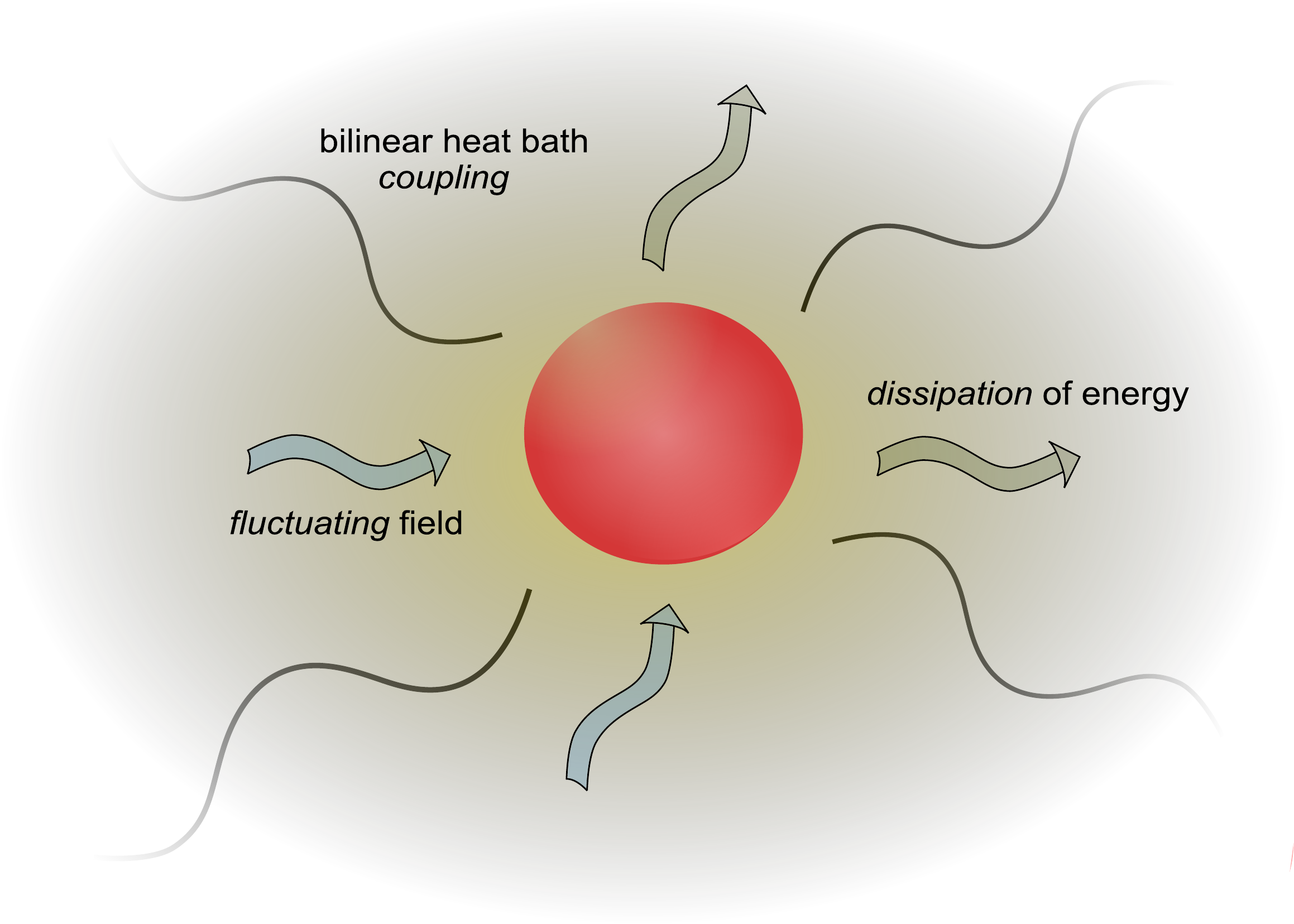}
\caption{\label{opsys} Illustration of a distinct quantum system embedded in a thermal reservoir. The coupling to the environment provokes phenomena such as fluctuating forces that act on the system and the loss of energy due to dissipation.}
\end{center}
\end{figure}

Within the quantum regime the dynamical properties of the reduced
density matrix are paramount. By tracing out reservoir degrees of
freedom from the global dynamics, the focus is narrowed towards a
relevant subsystem. Any systematic treatment of dynamical features like decoherence of
quantum states, dissipation of energy, relaxation to equilibrium or
non-equilibrium steady states requires a consistent procedure to
distill a dynamical map or an equation of motion for the reduced
density matrix from the unitary evolution of system and reservoir.

The Markovian approximation typically generalizes the classical
probabilistic technique of a dynamical semigroup in analogy to the
differential Chapman-Kolmogorov equation \cite{cgard09}.
The finite-dimensional mathematical framework of quantum dynamical
semigroups traces back to the seminal work by Gorini, Kossakowski,
Sudarshan \cite{kossa7201, kossa7202, gorin76} and simultaneously by
Lindblad \cite{lindb76}. While the resulting quantum master equations
of Lindblad form provide an easy-to-use set of tools for many
applications, their perturbative nature fails in the presence of
strong environment coupling, long correlation time scales or
entanglement in the initial state. The degree of non-Markovianity
which is inherent to the density matrix evolution and which causes
pronounced retardation effects in reservoir-mediated self-interactions
constitutes an active field of research \cite{brlapi09, chman14,
  pimasu08, vamapabrpi11, wolf08}. Whereas considerable advances have
been made in the characterization of dynamical generators as
non-Markovian \cite{rihupl14}, much less attention is paid to the question
of how non-Markovian behavior arises from the Hamiltonian description
of a system-reservoir model.

Beyond perturbative dynamics of memoryless master equations and
related strategies like quantum jumps \cite{hekpek13} or quantum state
diffusion \cite{dalcasmo92, duzori92, brpe95, giper92, giper98},
a pool of numerically exact simulation methods has been developed, each
with specific strengths and specific weaknesses.
One can distinguish between methods set-up in the full Hilbert space of system and bath degrees of freedom
and those considering the dynamics of the reduced density operator of the system alone.
The former include approaches based on e.g.\ the Numerical Renormalization Group (NRG) \cite{nrg},
the Multiconfiguration Hartree (MCTDH) \cite{mctdh}, and the Density Matrix Renormalization Group (DMRG)
\cite{prchhupl10}. The latter can all be derived from the path integral formulation pioneered by
Feynman and Vernon \cite{feve63,grscin88,weiss12}. They treat the functional integration either directly such
as the Path Integral Quantum Monte Carlo (PIMC) \cite{egma94, muehl05,kast13}
and the Quasi-Adiabatic Propagator (QUAPI) \cite{mamak94} or cast it in some form of time evolution equations. This is by no means straightforward
due to the bath induced time retardation, a problem that always appears at lower temperatures.
Equivalence to the path integral expression is then only guaranteed for
 a nested hierarchy of those equations \cite{ista05} or time evolution equations
 carrying stochastic forces \cite{strunz,stgr02,shao} from
 which the reduced density follows after a proper averaging.

These stochastic approaches exploit the intimate connection between
 the description of a quantum reservoir in terms of an influence
functional and stochastic processes. In fact, influence functionals do not only arise when a
partial trace is taken over environmental degrees of freedom, they are
also representations of random forces sampled from a classical
probability space \cite{feve63}. This stochastic construction can be
reversed, leading to an unraveling of quantum mechanical influence
functionals into time-local stochastic action terms \cite{stgr02}; we
thus obtain the dynamics of the reduced system through statistical
averaging of random state samples generated by numerically solving a single
time-local stochastic Liouville-von Neumann equation (SLN \cite{st04,
  schmidt2011, imohko15}). Compared to other methods
  this provides a very transparent formulation of non-Makrovian
  quantum dynamics with the particular benefit that the consistent
  inclusion of external time dependent fields is straightforward \cite{schmidt2011}.

Since the random forces
resulting from the exact mapping of the quantum reservoir to a probability space
are not purely real, the resulting
stochastic propagation is non-unitary. Therefore, the signal-to-noise
ratio of empirical statistics based on SLN propagation deteriorates
for long-time. This issue is reminiscent of the sign problem in
real-time path integrals \cite{macha90} and represents a major hurdle
in solving the most general c-number noise stochastic Liouville-von
Neumann equation for time intervals much longer than the timescales of
relaxation and dephasing.

Here we modify a strategy, recently presented by one of us
\cite{stock16}, which uses a projection operator \cite{naka58, zwan60}
based finite-memory scheme to solve the complex-noise SLN numerically.
Finite-memory stochastic propagation (FMSP) significantly lowers the
effect of statistical fluctuations and leads to a much faster
convergence of long-time sample trajectories, with a gain in
computational efficiency by several orders of magnitude.
While the relevant memory timescales used before were reservoir
correlation times \cite{stock16}, we use a different projector here,
adapted to the case of finite dephasing times. If either the
dissipative coupling or the reservoir temperature exceed certain
thresholds, this results in a shorter memory time and better
statistics.

We apply this framework to two- and three-level systems and compare
numerical data with perturbative predictions. Particular emphasis is put on the transition from coherent to
incoherent population dynamics. Since the new method allows for converged
simulations also in the long time domain, transfer rates for sequential hopping
and super-exchange \cite{weiss12} can be extracted which are of relevance for charge or energy transfer
in molecular aggregates and arrays of artificial atoms, e.g. quantum dot structures.

The paper is organized as follows: We start in Sec.~\ref{sln} with a concise discussion of the SLN and
its simplified version for ohmic spectral densities. The new scheme based in projection operator techniques
is introduced in Sec.~\ref{tcbd}, before the two-level system in Sec.~\ref{twolevel} and the three-level structure
in Sec.~\ref{threelevel}  are analyzed.

\section{Stochastic Representation Of Open System Dynamics}\label{sln}
We consider a distinguished system which is embedded in an environment
with a large number of degrees of freedom. The Hamiltonian of such a
model comprises a system, a reservoir and an interaction term
\begin{equation}
H = H_S + H_I + H_R.
\end{equation}
For bosonic elementary excitations of the reservoir we assume
$H_R=\sum_k\hbar\omega_k b^{\dagger}_k b_k$ together with a bilinear
coupling part $H_I=q\cdot\mathcal E$ that links the system coordinate
$q$ to the bath force $\mathcal E = \sum_k c_k
(b_k^{\dagger}+b_k)$. From the unitary time evolution of the global
density matrix $W$ that belongs to the product space $\mathcal
H = \mathcal H_S \otimes \mathcal H_R$, we recover the reduced density
matrix $\rho$ by a partial trace over the reservoir's degrees of
freedom
\begin{equation}
\rho(t)=Tr_R\lbrace \pazocal U(t,t_0)W(t_0)\pazocal U (t,t_0)\rbrace
\end{equation}
and a factorizing initial condition
$W(t_0)=\rho(t_0)\otimes\rho_R$. We thereby assume the reservoir to be
initially in thermal equilibrium, $\rho_R = Z_R^{-1}e^{-\beta H_R}$.

While traditional open system techniques focus on a perturbative
treatment in the interaction Hamiltonian $H_I$, we derive our
stochastic approach from influence functionals, a path integral
concept introduced by Feynman and Vernon \cite{feve63}, applying to
reservoirs with Gaussian fluctuations of the force field $\mathcal
E(t)$,
\begin{equation}
\int\pazocal D q (\tau)\pazocal D q' (\tau') \pazocal A[q(\tau)]\pazocal A^{\ast}[q'(\tau')] \pazocal F [q(\tau)q'(\tau')].
\label{FVdoubint}
\end{equation}
Here $\pazocal A$ is the probability amplitude for paths governed by
the system action alone (a pure phase factor), and
\begin{eqnarray}
\ln \pazocal F [q(\tau)q'(\tau')] &=& - \frac{1}{\hbar^2}
\int_0^t d\tau \int_0^\tau d\tau'
(q(\tau)-q'(\tau))\nonumber\\
&\times&[L(\tau-\tau')q(\tau') - L^*(\tau-\tau')q'(\tau')].
\label{eq:lnF}
\end{eqnarray}
All effects of the dissipative environment on the propagation of
the system can thus be expressed in terms of the \emph{free}
fluctuations of the reservoir, characterized by the  complex
force correlation function
\begin{eqnarray}
L(\tau)&\equiv&\langle\mathcal E (\tau)\mathcal E (0)\rangle_R =
L'(\tau)+iL''(\tau) \nonumber
\\ &=&\frac{\hbar}{\pi}\int_0^{\infty}d\omega\:
J(\omega)\frac{\cosh[\omega(\hbar\beta/2-it)]}{\sinh(\hbar\beta\omega/2)}
.
\label{Lcorr}
\end{eqnarray}
Here we have introduced a spectral density $J(\omega)$, which can be
determined microscopically from the frequencies and couplings of a
quasi-continuum of reservoir modes, or, alternatively, from the
Fourier transform of the imaginary part of eq. (\ref{Lcorr}) when the
correlation function has been obtained by other means. The latter
definition is somewhat more general than the model of an oscillator
bath. It is to be noted that the thermal timescale $\hbar\beta$ must
be considered long in a quantum system with thermal energy $k_B
T\equiv 1/\beta$ smaller than the level spacing. In the opposite limit
$\hbar\beta \rightarrow 0$ the classical version of the
Fluctuation-Dissipation theorem \cite{cawe51} can be recovered from
eq. (\ref{Lcorr}).

Due to the non-local nature of the influence functional
(\ref{eq:lnF}), there is no simple way to recover an equation of
motion from eq. (\ref{FVdoubint}). However, since $\pazocal F
[q(\tau)q'(\tau')]$ is a Gaussian functional of the path variables, it
shows great formal similarity to generating functionals of classical
Gaussian noise.  A classical process formally equivalent to a quantum
reservoir can thus be constructed by means of a stochastic
decomposition \cite{cald83, stgr02}. The resulting stochastic
Liouville-von Neumann equation (SLN) contains two stochastic processes
$\xi(t)$ and $\nu(t)$, corresponding to two independent functions of
the functional $\pazocal F$,
\begin{eqnarray}
\frac{d}{dt}\rho_{z}(t)&=&-\frac{i}{\hbar}[H_S,\rho_{z}(t)]\nonumber \\
&+&\frac{i}{\hbar}\xi(t)[q,\rho_{z}(t)]+\frac{i}{2}\nu(t)\{q,\rho_{z}(t)\}.
\label{SLN}
\end{eqnarray}
We thus map the reduced system evolution to stochastic propagation
in probability space of Gaussian noise forces with zero bias and
correlations which match the quantum mechanical correlation function
$L(t-t')$,
\begin{eqnarray}
\langle\xi(t)\xi(t')\rangle_R&=& L'(t-t'),\label{corrxixi}\\
\langle\xi(t)\nu(t')\rangle_R&=&(2i/\hbar)\Theta(t-t') L''(t-t'),
\label{corrxinu}\\
\langle\nu(t)\nu(t')\rangle_R&=&0.\label{corrnunu}
\end{eqnarray}
It is obvious that there are no real-valued processes with these
correlations, however, complex-valued stochastic processes which obey
eqs. (\ref{corrxixi})--(\ref{corrnunu}) do exist.

Even though eq. (\ref{SLN}) contains no term recognizable as a damping
term (in a mathematical sense), yet it provides an exact numerical
approach to open quantum systems with any type of Gaussian
reservoir: Stochastic samples $\rho_z$ obtained by propagating
(\ref{SLN}) with specific noise samples have no obvious physical
meaning. The \emph{physical} density matrix, whose evolution is
damped, is obtained by averaging the samples, $\rho=\mathbbm M \left[
  \rho_z\right]$.

There is also a version of the SLN which is designed for reservoirs with ohmic characteristic,
i.e.\ for spectral densities of the form $J(\omega)\sim \omega$ up to a high frequency cutoff $\omega_c$
being significantly larger than any other frequency of the problem (including the thermal time
$\hbar\beta$). This class of reservoirs is of particular relevance due to numerous realizations ranging from
atomic to condensed matter physics. In this case, the imaginary part of the reservoir correlation
function can be considered as the time derivative of a Dirac $\delta-$function,
\begin{equation}
L''(\tau) = \frac{\eta}{2} \frac{d}{d\tau} \delta(\tau)\, .
\end{equation}
Accordingly, memory effects arise only from the real part $L'(\tau)$, while
the imaginary part can be represented by a time-local damping operator
acting on the reduced density matrix. This results in the simplified so-called SLED dynamics (stochastic Liouville
equation with dissipation) \cite{StockburgerMak1999} with one real-valued
noise force $\xi(t)$,
\begin{eqnarray}
\frac{d}{dt}\rho_{\xi }&=&\frac{1}{i\hbar}\left([H_S,\rho_{\xi}]-\xi(t)[q,\rho_{\xi}]\right)\nonumber\\
&+&\frac{\gamma}{2i\hbar}[q,\{p,\rho_{\xi}\}]
\label{SLED}
\end{eqnarray}
where $\gamma = \eta/m$.
Eq. (\ref{SLED}) has been derived for potential models with canonical
variables $q$ and $p$ with $[q,p] = i\hbar$. In a wider context, it is
still valid for moderate dissipative strength
\cite{Stockburger1999,jsto06} with the substitutions $m \to 1$ and $p
\to (i/\hbar) [H,q]$.

Both stochastic formulations for open quantum dynamics
eq. (\ref{SLN}) and eq. (\ref{SLED}) have
been successfully used to solve problems in a variety of areas. They apply to systems with discrete
Hilbert space as well as continuous degrees of freedom and have the particular benefit
that they allow for a natural inclusion of external time dependent fields irrespective of
amplitude and frequency. Specific applications comprise spin-boson dynamics \cite{st04}, optimal
control of open systems \cite{schmidt2011}, semiclassical dynamics \cite{kogrostank08}, molecular energy transfer \cite{imohko15}, generation of entanglement \cite{scstank13},
and  heat and work fluctuations \cite{sccapesuank15, stmotz16}, to name but a few.

\section{Time-Correlated Blip Dynamics}\label{tcbd}

However, there is a price to be paid for the generality and simplicity
of eqs. (\ref{SLN}), (\ref{SLED}). The correlation function eq. (\ref{corrnunu}) requires the complex-valued
process $\nu(t)$ to have a random phase; hence eq. (\ref{SLN})
describes non-unitary propagation. As in the paradigmatic case of
multiplicative noise, geometric Brownian motion \cite{cgard09}, the
stochastic variance of observables [taken with respect to the
probability measure of $\xi(t)$ and $\nu(t)$] grows rapidly with
increasing time $t$, making the computational approach prohibitively
expensive in the limit of very long times.

This problem has recently been solved by one of us \cite{stock16} by
formally identifying the stochastic averaging as a projection
operation. This allows the identification of relevant and irrelevant
projections of the ensemble of state samples, with beneficial
simplifications for finite memory times of the reservoir. The
finite-memory stochastic propagation (FMSP) approach has solved the
problem of deteriorating long-term statistics. However, the finite
asymptotic value of the sampling variance can still be high in the
case of strong coupling.  Here, we address this case
using a different projection operator, based on finite decoherence
timescales instead of reservoir correlation times. We restrict ourselves
to reservoirs with ohmic-type spectral densities for which the SLED in eq.~(\ref{SLED})
is the appropriate starting point; generalizations will be presented elsewhere.

In this section we first describe the general strategy and will then turn to
specific applications in the remainder. Since the latter focus on systems with discrete
Hilbert space, for convenience we make use of the language developed within
the path integral description of discrete open quantum systems.
There,  it is customary to label path segments where the path labels $q(\tau)$
and $q'(\tau)$ differ as ``blips'', contributing to off-diagonal
matrix elements (coherences) of $\rho$ and intervening periods with
equal labels as ``sojourns'', contributing to diagonal elements.

In an open quantum system, dephasing by the environment sets an
effective upper limit on the duration of a blip. This observation can
be transferred to our stochastic propagation methods using projection
operators. The operator
\begin{equation}\small
{\mathcal P}: \rho \to \left(
\begin{array}{cccc}
\rho_{11} &0 & \cdots& 0\\
0 & \rho_{22} & \ddots & \vdots\\
\vdots & \ddots & \ddots & 0 \\
0 & \cdots & 0 & \rho_{nn}
\end{array}
\right)
\end{equation}
projects on sojourn-type intermediate states, while its complement
${\mathcal Q} = {\mathbbm 1} - {\mathcal P} $ projects on blip-type
states.

With these projectors, eq. (\ref{SLED}) can be rewritten as a set of
coupled equations of motion for diagonal and off-diagonal elements of
$\rho_\xi$,
\begin{equation}
\small
\frac{d}{dt}
\begin{pmatrix}
	\rho_{\mathcal P} \\
	\rho_{\mathcal Q}
\end{pmatrix} = \begin{pmatrix}
	0 && \mathcal P \pazocal L_{det} \mathcal Q \\
	\mathcal Q \pazocal L_{det} \mathcal P && \mathcal Q [\pazocal L_{det}+\pazocal L_{\xi}]\mathcal Q
\end{pmatrix}
\begin{pmatrix}
	\rho_{\mathcal P} \\
	\rho_{\mathcal Q}
\end{pmatrix}
\label{sys3}
\end{equation}
with a deterministic Liouvillian superoperator $\pazocal L_{det}
\equiv \frac{1}{i\hbar}[H_S,\cdot]+\frac{\gamma}{2 i \hbar}
       [q,\{p,\cdot\}]$ and a stochastic superoperator $\pazocal
       L_{\xi} \equiv \frac{i}{\hbar} [q,\cdot]\xi(t)$.  The
       propagation of the SLED equation (\ref{SLED}) can now be
       re-cast in the form of a Nakajima-Zwanzig equation
       \cite{naka58,zwan60}
\begin{eqnarray}
&& \frac{d}{dt} \rho_{\mathcal P}  = {\mathcal P}{\cal L}  \rho_{\mathcal P}  \nonumber\\
&& + {\mathcal P}{\cal L} \int\limits_{t^*}^t dt'
\exp_>\left(\int\limits_{t'}^t ds{\mathcal Q}{\mathcal L}(s)\right){\mathcal Q}{\cal L}  \rho_{\mathcal P} (t') ,
\label{eq:zwanzig}
\end{eqnarray}
where $\rho_{\mathcal P}$ now is the projected (diagonal) part of the
density matrix, ${\mathcal L}\rho$ is equal to the r.h.s. of
eq. (\ref{SLED}), and the symbol ``$>$'' denotes time ordering.

The time-ordered exponential now represents propagation during
``blip'' periods. Generally, the lower integration boundary $t^*$
needs to be set to zero if full equivalence to equations of type
(\ref{sys3}) is required. However, when it is known that dephasing
sets an upper limit to blip times, $t^*$ can be raised, provided that
$t-t^*$ remains large compared to the dephasing time.

In order to translate this approach into an algorithm, it is
advantageous not to choose $t-t^*$ constant: The number of different
initial conditions for the irrelevant part must be kept manageable.

\begin{figure*}
\begin{center}
\includegraphics[width = 16 cm]{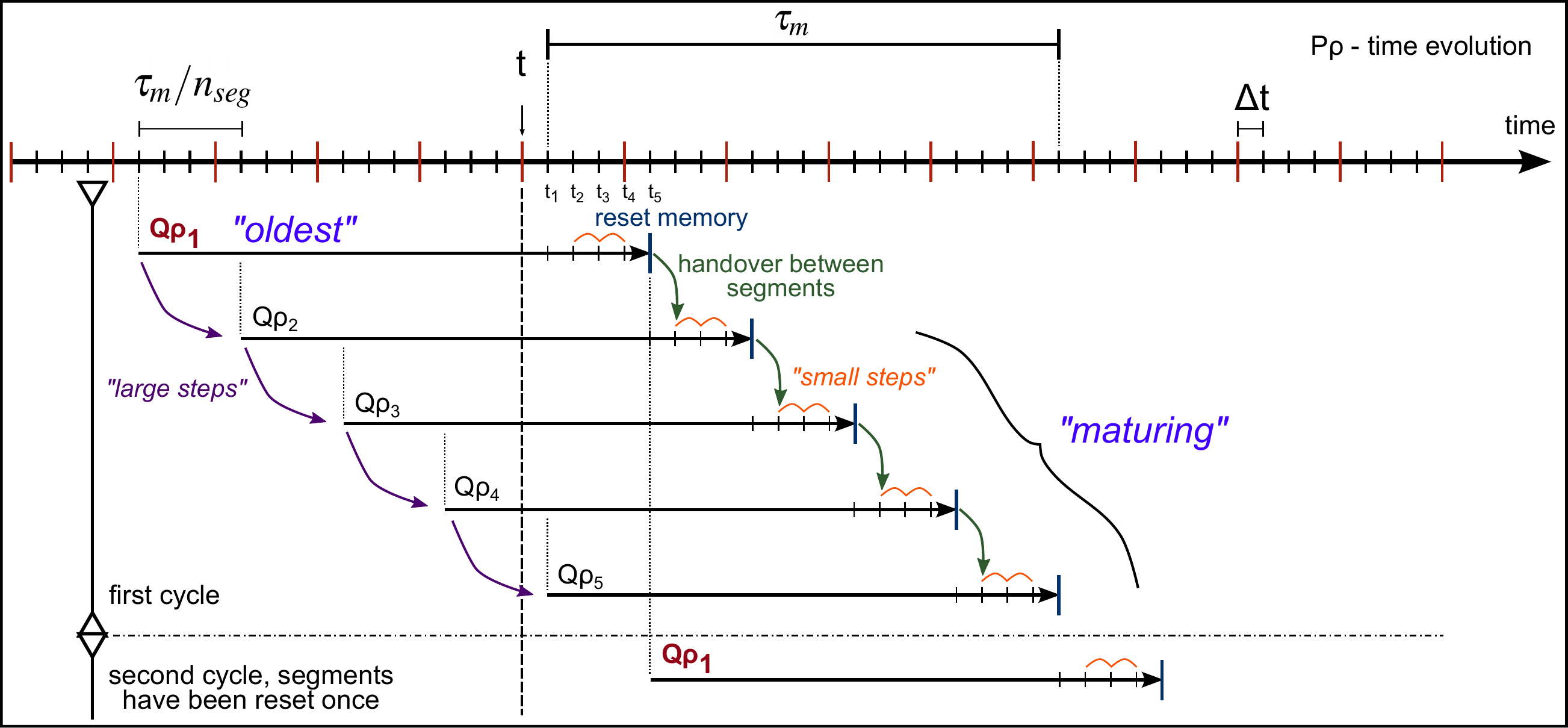}%
\caption{\label{fig1}Scheme of "memory recycling" in the Time-Correlated Blip Dynamics (TCBD).}
\end{center}
\end{figure*}
% It
%is more practical to raise $t^*$ in steps larger than the numerical
%propagation timestep, but smaller than the dephasing time.

Fig. \ref{fig1} provides an illustration of the resulting algorithm,
Time-Correlated Blip Dynamics (TCBD). The propagation of coherences is
realized in multiple overlapping segments on the time axis.  An
arbitrary, but fixed number of segments $n_{seg}$, i.e. $\mathcal Q
\rho_j$ with $j \in \{1,2,3,\ldots, n_{seg}\}$ is used throughout the
propagation. These segments are initialized with staggered starting
times.  With an inter-segment spacing $\sim \tau_m/n_{seg}$ chosen
larger than the numerical timestep $\Delta t$. The individual segments
comprise a maximum memory window of $\tau_m$. For each propagation step at a given time $t$, 
the segment with the longest history among all $\mathcal Q
\rho_j$ is used in the propagation of $\mathcal P \rho$. Whenever one
segment has ``aged'' beyond $\tau_m$, it is reset and replaced by its
next best follower, thus starting a continuous recycling of memory
trails.

Formally limiting the maximum blip length leads to improved statistics
of our stochastic simulations, since the stochastic part of ${\mathcal
  L}$ applies only to blip periods. The choice of a pre-defined number
of memory segments (independent of the propagation timestep) reduces
the required numerical operations significantly. Compared to a naive
solution of eq. (\ref{eq:zwanzig}) with fixed difference $t-t^*$,
resulting in an unfavorable scaling of complexity with $\Delta t$ as
$\mathcal O_{\text{naive}} \left(\left[\frac{t}{\Delta
    t}\right]^2\right)$, we lower the complexity to $\mathcal
O_{\text{TCBD}} \left(\frac{t}{\Delta t}\cdot n_{seg}\right)$,
typically an order of magnitude lower in absolute terms.

In the remainder of this work, we apply the new non-Markovian
propagation method TCBD to both a two-level system (TLS) immersed in
a heat bath (spin-boson model) and a quantum system that is
effectively restricted by a three-dimensional Hilbert space. In the
transparent context of these discrete systems, we compare equilibrium
properties to an analytical theory of dissipative two-state evolution,
the non-interacting blip approximation (NIBA) and uncover
super-exchange phenomena due to virtual particle transfer.

\section{Spin-Boson model}\label{twolevel}

The spin-boson model is a generic two-state system coupled linearly to
a dissipative environment
\begin{equation}
H_S = \frac{\hbar \epsilon}{2}\sigma_z - \frac{\hbar\Delta}{2}\sigma_x .
\label{sbh}
\end{equation}
The coupling to the environment is conventionally taken as $H_I =
\sigma_z \cdot\mathcal E$. The ohmic environment is described by
a spectral density of the generic form
\begin{equation}
J(\omega)=\frac{\eta\omega}{(1+\omega^2/\omega^2_c)^2}
\label{osd}
\end{equation}
where the constant $\eta$ denotes a coupling constant,
and $\omega_c$ is a UV cutoff. 
In the context of the spin-boson problem
one conventionally works with the so-called Kondo parameter $K= 2\eta
/ (\pi\hbar)$ as a dimensionless coupling parameter.
The SLED and the TCBD are then applicable for moderate damping $K< 1/2$. Subsequently, 
one sets $q=\sigma_z$, $m=1$ and $p = - \Delta \sigma_y$ so that the
deterministic and stochastic superoperators introduced in
eq. (\ref{sys3}) read
\begin{eqnarray}
\pazocal L_{det} &=& \frac{1}{i\hbar}[H_S,\cdot]-\frac{\gamma}{2 i
  \hbar} \Delta[\sigma_z,\{\sigma_y,\cdot\}]\nonumber \\ \pazocal
L_{\xi} &=& \frac{i}{\hbar}[\sigma_z,\cdot]\xi(t).
\end{eqnarray}
Parameterizing the reduced density matrix through pseudospin expectation values leads to an intuitive picture of the dynamics
\begin{equation}
\rho = \frac{1}{2}\left(
\begin{array}{cc}
1 + \langle\sigma_z\rangle & \langle\sigma_x\rangle -
i\langle\sigma_y\rangle \\
\langle\sigma_x\rangle + i\langle\sigma_y\rangle & 1 - \langle\sigma_z\rangle
\end{array}
\right).
\end{equation}
We will later make use of the fact that the diagonal part is
determined by the single parameter $\langle\sigma_z\rangle$.

It is instructive to compare our projection technique with the
analytic NIBA approach \cite{legcha87}. Path integral techniques were
first used to derive the NIBA; with discrete path variables $\sigma =
\pm 1$, eq. (\ref{FVdoubint}) reads
\begin{equation}
\int\pazocal D \sigma (\tau)\pazocal D \sigma' (\tau') \pazocal A[\sigma(\tau)]\pazocal A^{\ast}[\sigma'(\tau')] \pazocal F [\sigma(\tau)\sigma'(\tau')].
\label{nidoubint}
\end{equation}
Since the path functions $\sigma(\tau)$ are piecewise constant, their
derivatives are sums of delta functions at isolated
points. Eq. (\ref{eq:lnF}) can then be integrated by parts; this results
in a double sum over ``interactions'' between discrete ``charges''
(jumps in path variables), which depend on the twice-integrated
reservoir correlation function
\begin{eqnarray}
Q(t) &=& \frac{\hbar}{\pi}\int_0^{\infty} d\omega \frac{J(\omega)}{\omega^2}\{ \coth(\frac{\hbar\beta\omega}{2})[1-\cos(\omega t)] \nonumber\\
&& + i \sin(\omega t)\}.
\label{qfun}
\end{eqnarray}
The NIBA approach assumes that the resulting form of eq. (\ref{eq:lnF})
can be approximated by omitting interactions between ``charges'' which
are not part of the same blip. The Laplace transform of the path sum
can then be given in analytic form.

For the present purpose, it is more instructive to note that this result
can also be cast in the form of an equation of motion \cite{Dekker1987}
\begin{equation}
\frac{d\left\langle \sigma_z\right\rangle_t}{dt}=\int_0^{t}dt'\left[K_z^{(a)}(t-t')-K_z^{(s)}(t-t')\left\langle \sigma_z\right\rangle_{t'}\right]
\label{NIBA}
\end{equation}
with integral kernels that depend on the relative time $\tau = t-t'$
\begin{eqnarray}
K_z^{(a)}&=&\Delta^2 \sin(\epsilon \tau) e^{-Q'(\tau)}\sin(Q''(\tau))\\
K_z^{(s)}&=&\Delta^2 \cos(\epsilon\tau)e^{-Q'(\tau)} \cos(Q''(\tau)).
\label{difac}
\end{eqnarray}
The function $Q(\tau)$ is given by
\begin{eqnarray}
Q'(\tau)&=& 2 K \ln\left(\frac{\hbar\beta\omega_c}{\pi}\sinh \frac{\pi
  \tau}{\hbar \beta}\right) \\ Q''(\tau)&=& 2 K
\label{disfa2}
\end{eqnarray}
for large $\omega_c$ and moderate damping ($K<1/2$).

The integro-differential equation (\ref{NIBA}) has a mathematical
structure similar to the Nakajima-Zwanzig equation
(\ref{eq:zwanzig}). This is not a coincidence. The dynamical variable
$\langle\sigma_z\rangle$ of eq. (\ref{NIBA}) is representative of the
entire diagonal part of the density matrix, after averaging over
reservoir degrees of freedom (or, in our case, noise representing the
reservoir). NIBA thus is related to the application of a different
projection operator $\bar{\mathcal P}$ which sets off-diagonal
elements to zero \emph{and} takes the expectation value of the
diagonal elements (the relation $\bar{\mathcal P}^2 = \bar{\mathcal
  P}$ is obvious).

However, the integral kernels of NIBA are equivalent to propagating
with ${\mathcal Q} {\mathcal L}$ rather than $\bar{\mathcal Q}
{\mathcal L}$. Therefore eq. (\ref{NIBA}) is not quite a
Nakajima-Zwanzig equation. The difference between NIBA and TCBD could
thus succinctly be stated in the following manner: NIBA tacitly omits
the projection $({\mathcal P}-\bar{\mathcal P})\rho$ from the dynamics.

Apart from providing reference data and being a useful conceptual
reference point, the NIBA theory provides us with a quantitative model
for the system's decoherence time based on the reservoir's dissipative
properties. When the function $Q'(\tau)$ increases with $\tau$ at long
times, long blips are suppressed as the dissipative factor
$e^{-Q'(\tau)}$ in eq. (\ref{difac}) becomes smaller. An estimate for
a memory window $\tau_m$ of the TCBD's $\rho_Q$ segments can be
obtained for $\tau \gg \hbar\beta$ and sufficiently strong suppression
of intra-blip interactions $e^{-Q'(\tau)}\ll 1$. Considerably small
errors are achieved by memory lengths of the order of
\begin{equation}
\tau_m \sim \frac{4\hbar\beta}{2K\pi}.
\end{equation}
Longer memory time frames $\tau_m$ reduce potential errors due to a truncation of the off-diagonal trajectories. At the same time, they increase the amount of random noise that accumulates along the stochastic propagation of $\rho_Q$ and smears out the original transfer signal. While the TCBD method allows to manually access intra-blip correlation lengths through $\tau_m$, it puts no restrictions on inter-blip interactions, thereby extending the NIBA theory.
%\begin{figure}[h]
%\begin{center}
%\includegraphics[width = 9 cm]{noisecomp1.png}%
%\caption{\label{fig2} Comparison of TCBD and SLED expectation values $\left\langle \sigma_z\right\rangle_t$ within the regime of $K\approx 0.24$, $\Delta=1$, $\epsilon=0$, $\omega_c=10$, $\tau_m =2$ and $n_{samp}=2500$.}
%\end{center}
%\end{figure}

\begin{figure}[h]
\begin{center}
\includegraphics[width = 9 cm]{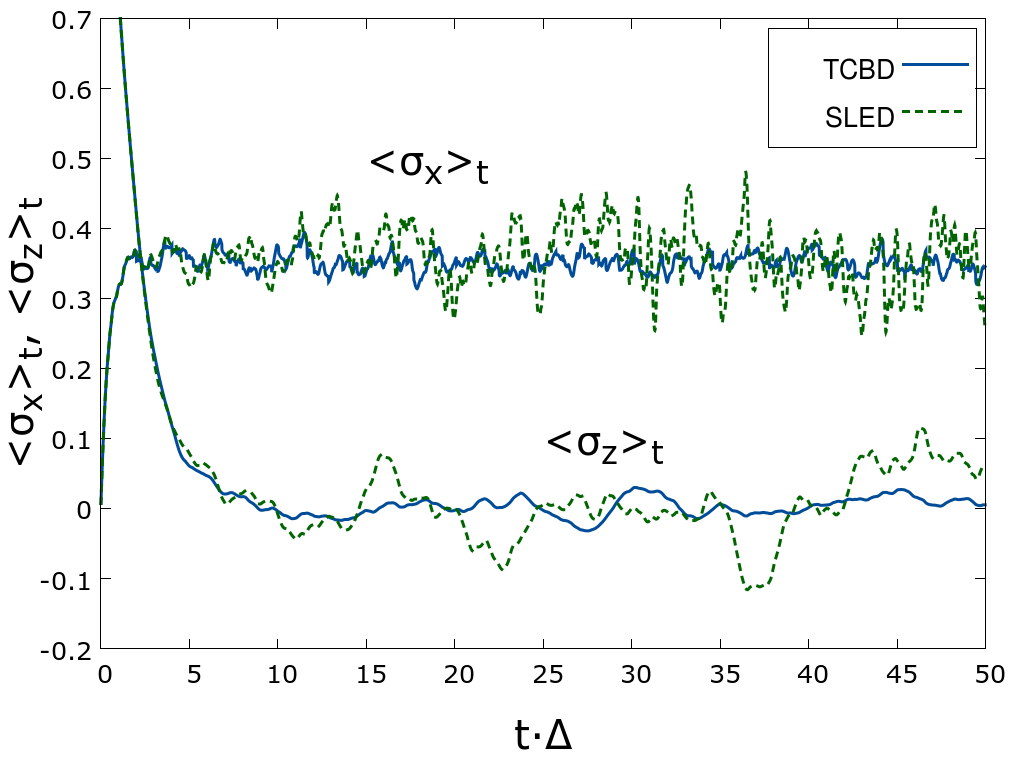}%
\caption{\label{fig2} Comparison of the new TCBD and the full SLED [eq.~(\ref{SLED})] for expectation values $\left\langle \sigma_x\right\rangle_t$ and $\left\langle \sigma_z\right\rangle_t$ of a spin-boson model with $\epsilon=0$, $\beta = 0.7$, $K = 0.24$, $\omega_c=10$, $\tau_m =2$ and $n_{samp}=2500$; frequencies in units of $\Delta$.}
\end{center}
\end{figure}

\begin{figure}[h]
\begin{center}
\includegraphics[width = 9 cm]{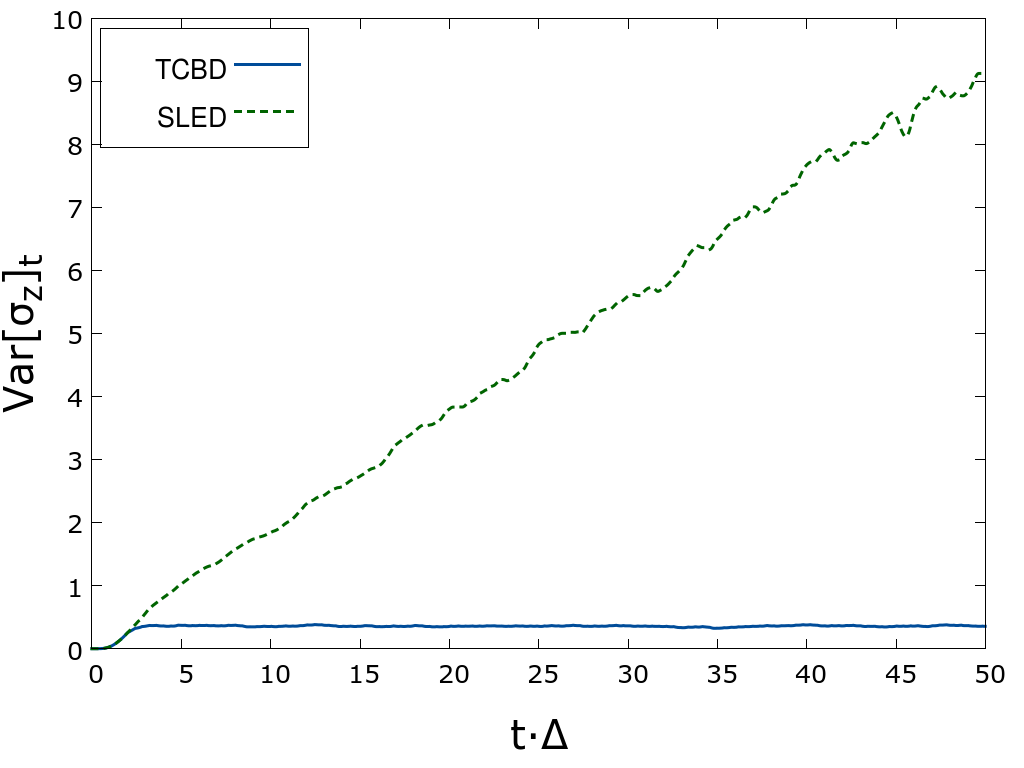}%
\caption{\label{fig3} Comparison of the numerical variance for TCBD and SLED for $\left\langle \sigma_z\right\rangle_t$; parameters are as in Fig.~\ref{fig2}.}
\end{center}
\end{figure}

The efficiency gain of a numerical spin-boson simulation based on the SLED eq. (\ref{SLED}), and the proposed TCBD method, proves to be especially striking in the case of strong dephasing, i.e. for large coupling parameters $K$. Beside the dynamical observables $\langle\sigma_x\rangle_t$ and $\langle\sigma_z\rangle_t$, Fig. \ref{fig2} and Fig. \ref{fig3} compare the variance of the stochastic sample trajectories $\text{Var}[\sigma_z]_t$ of a straightforward SLED solution to the TCBD propagation of the reduced system.  Apparently, both first and second order statistics confirm substantial faster convergence of the TCBD scheme to the thermal equilibrium $\langle\sigma_z\rangle_{\infty}=0$ as well as a significant reduction in sample noise.\\

Despite its sound quantitative applicability in a wide range of fields, the NIBA flaws in describing long time dynamics of $\langle\sigma_j\rangle_t$ ($j=x,y,z$) correctly at low temperatures $T$ and finite energy bias $\epsilon$. Considering the equilibrium values for the $\langle\sigma_z\rangle_t$ component \cite{weiss12}, the NIBA result comes down to a loss in symmetry and strict localization for zero temperature
\begin{equation}
\left\langle \sigma_z\right\rangle_{\infty}^{NIBA}=\tanh\left(\frac{\hbar\epsilon}{2 k_B T}\right)\stackrel{T\rightarrow 0}{\longrightarrow}\text{sgn} (\epsilon).
\end{equation}
This confinement to one of the wells stands in contrast to the weak-damping equilibrium with respect to thermally occupied eigenstates of the TLS
\begin{equation}
\left\langle \sigma_z\right\rangle_{\infty}^{eff}=\frac{\epsilon}{\Delta_{eff}}\tanh\left(\frac{\hbar\Delta_{eff}}{2 k_B T}\right)\stackrel{T\rightarrow 0}{\longrightarrow}\frac{\epsilon}{\Delta_{eff}}
\end{equation}
with effective tunneling matrix element
\begin{equation}
\Delta_{eff}=\left[\Gamma(1-2K)\cos(\pi K)\right]^{1/2(1-K)}\left(\frac{\Delta}{\omega_c}\right)^{K/(1-K)}\Delta.
\label{tuneff}
\end{equation}
While the equilibrium prediction of the NIBA theory clearly fails in the presence of an even infinitesimal bias energy, the TCBD provides correct equilibration in the scaling limit. Note that due to the finite frequency cutoff in eq. (\ref{osd}), the TCBD approaches a steady state value determined by the effective tunnel splitting $\Delta_{eff}$ for coupling parameters $K<1$.\\

\begin{figure}[h]
\begin{center}
\includegraphics[width = 9 cm]{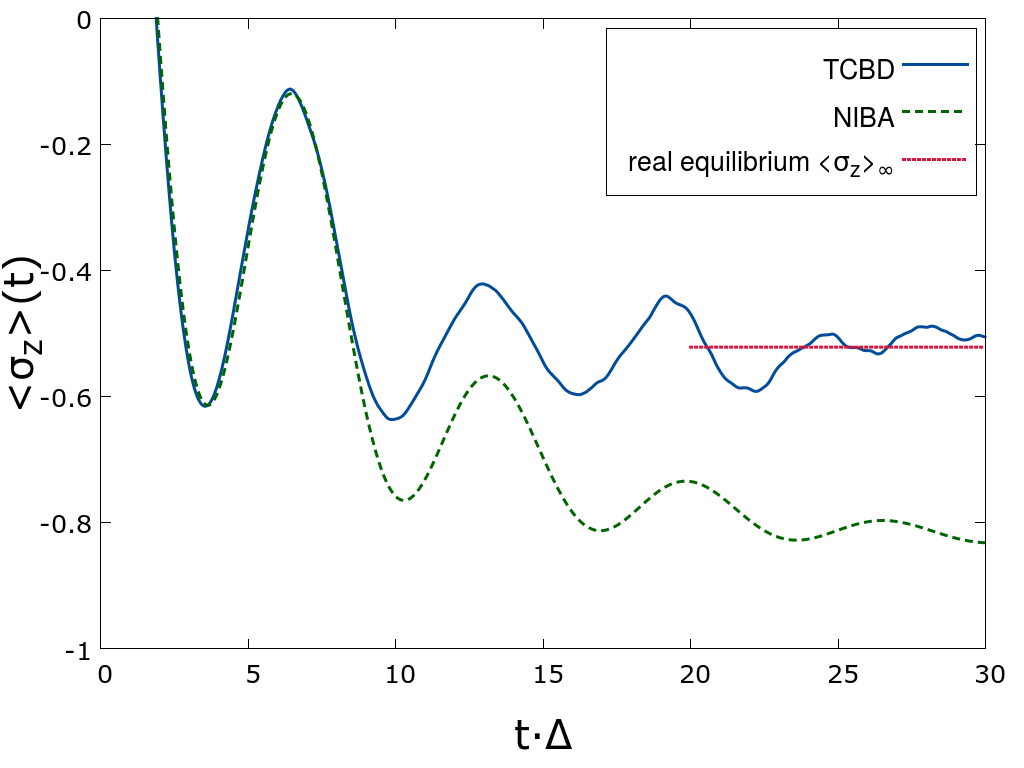}%
\caption{\label{fig4} NIBA and TCBD relaxation dynamics towards equilibrium of the population difference $\left\langle \sigma_z\right\rangle_t$ for $\epsilon=0.5$, $\beta=5$, $K = 0.1$, $\omega_c = 10$, $n_{samp}=100$ (in units of $\Delta$).}
\end{center}
\end{figure}

 Results in Fig. \ref{fig4} illustrate the relaxation process of  $\langle\sigma_z\rangle_t$ for a finite bias and low temperatures.
While both the NIBA and the TCBD provide nearly indistinguishable data up to times $t\cdot\Delta\approx 5$, significant deviations appear for longer times. We note in passing the numerical stability of the TCBD which allows to access also typical equilibration time scales.
 While in the TCBD approach all time-nonlocal correlations induced by the reservoir are consistently taken into account as a vital ingredient for the reduced dynamics, the NIBA neglects long-ranged interactions such as inter-blip correlations. In this sense, the TCBD method constitutes a systematic extension of the NIBA.

%-------------------------------------------------------------------------------------------------------------
%-------------------------------------------------------------------------------------------------------------
%-------------------------------------------------------------------------------------------------------------
%-------------------------------------------------------------------------------------------------------------
%-------------------------------------------------------------------------------------------------------------
%-------------------------------------------------------------------------------------------------------------

\section{Three-level system}\label{threelevel}

\begin{figure}[h]
\begin{center}
\includegraphics[width = 4 cm]{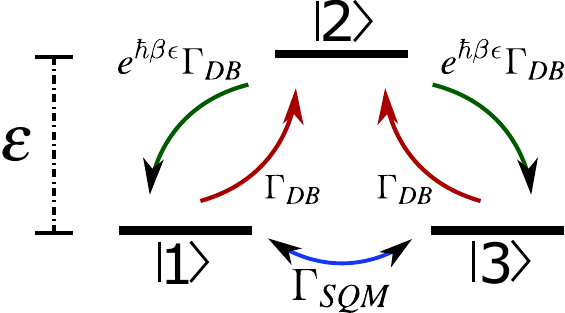}%
\caption{\label{fig5a} Symmetric three-state system donor-bridge-acceptor (DBA) with donor $\ket 1$ and acceptor $\ket 3$ states and a  bridge state $\ket 2$ elevated by an energy $\hbar\epsilon$. In case of incoherent population transfer, two transfer channels with transition rate $\Gamma_{DB}$ for sequential hopping and $\Gamma_{SQM}$ for super-exchange govern the dynamics.}
\end{center}
\end{figure}

We will now demonstrate the adaptability of the TCBD method for multilevel systems by investigating population transfer in a three-state structure. For this purpose, we consider (cf.~Fig. \ref{fig5a}) a symmetric donor-bridge-acceptor (DBA) system \cite{muean04, muehl05}, with degenerate donor state $\ket{1}$, acceptor state $\ket{3}$, and a bridge state $\ket{2}$ being energetically lifted by a bias of height $\epsilon$. Despite its simplicity the corresponding open quantum dynamics even in presence of an ohmic environment (\ref{osd}) is rather complex and has been explored in depth as a model to access fundamental processes such as coherent/incoherent dynamics and thermally activated sequential hopping versus long-range quantum tunneling (super-exchange).

Within the spin-1 basis $\{ \mathbbm 1, S_x, S_y, S_z \}$
%-----------------------------------------
%\begin{equation}
%S_x=\frac{1}{\sqrt{2}}\begin{pmatrix}
%	0 & 1 & 0 \\
%	1 & 0 & 1 \\
%	0 & 1 & 0
%\end{pmatrix},\:\:
%S_y=\frac{1}{\sqrt{2}i}\begin{pmatrix}
%	0 & 1 & 0 \\
%	-1 & 0 & 1 \\
%	0 & -1 & 0
%\end{pmatrix},\:\:
%S_z=\begin{pmatrix}
%	1 & 0 & 0 \\
%	0 & 0 & 0 \\
%	0 & 0 & -1
%\end{pmatrix}
%\end{equation}
%-----------------------------------------
the Hamiltonian of the system with site basis eigenvectors $\ket 1$, $\ket 2$ and $\ket 3$ reads
\begin{eqnarray}
H_S &=& \hbar \Delta S_x + \frac{\hbar\epsilon}{\sqrt{2}}(\mathbbm 1 - S_z^2) \nonumber\\
&=& \frac{\hbar}{\sqrt{2}} \begin{pmatrix}
	0 & \Delta & 0 \\
	\Delta & \epsilon & \Delta \\
	0 & \Delta & 0
\end{pmatrix}.
\label{3ham}
\end{eqnarray}
In eq. (\ref{SLED}), the position coordinate $q$ is then represented by the $S_z$, while the momentum operator $p$ follows from the respective Heisenberg equation of motion $\dot q = \frac{i}{\hbar}[H_S,q] = \Delta S_y$. Accordingly, $H_I = -S_z\cdot\mathcal E$ so that the deterministic and stochastic superoperators in eq. (\ref{sys3}) are derived as
\begin{eqnarray}
\pazocal L_{det} &=& \frac{1}{i\hbar}[H_S,\cdot]+\frac{\gamma}{2 i \hbar} \Delta[S_z,\{S_y,\cdot\}]\nonumber \\
\pazocal L_{\xi} &=& \frac{i}{\hbar} [S_z,\cdot]\xi(t).
\end{eqnarray}

\begin{figure}[h]
\begin{center}
\includegraphics[width = 9 cm]{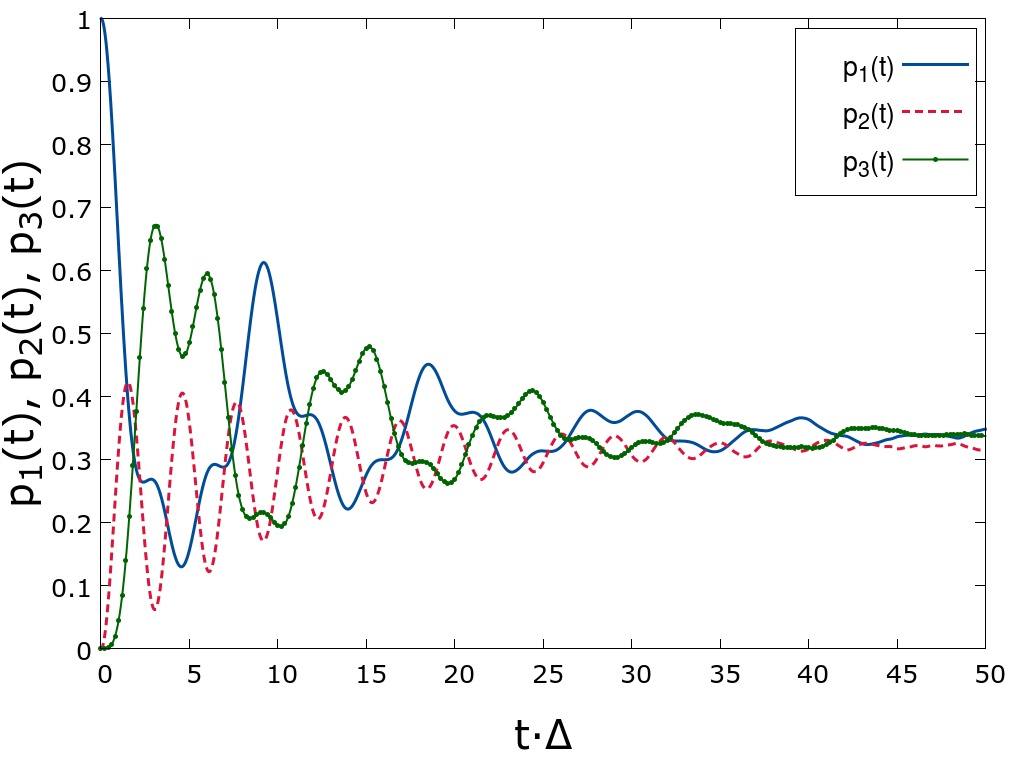}%
\caption{\label{fig5} Population dynamics $p_j(t)=\text{Tr}\{\ket j \bra j \rho(t)\}, j=\{1,2,3\}$ for $\epsilon = 1$ and $K = 0.24$, $\beta = 5$, $n_{samp}= 2\cdot 10^4$ and $\omega_c = 10$ (in units of $\Delta$).}
\end{center}
\end{figure}

\begin{figure}[h]
\begin{center}
\includegraphics[width = 9 cm]{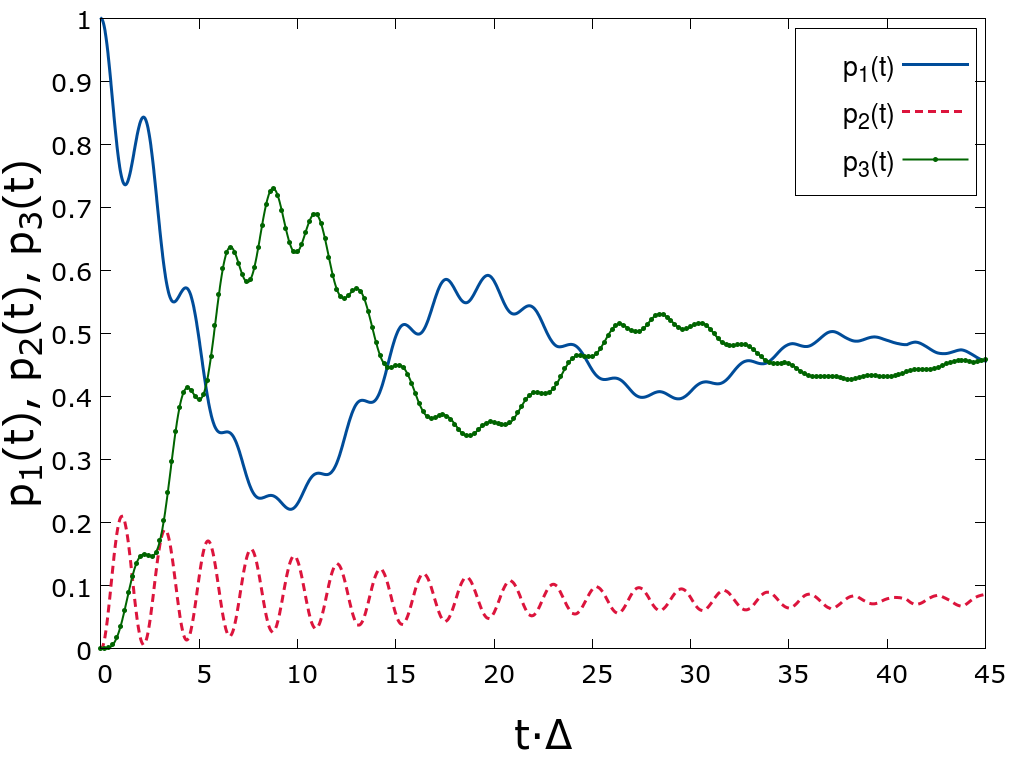}%
\caption{\label{fig6}  Same as in Fig.~\ref{fig5} but for  $\epsilon = 3$.}
\end{center}
\end{figure}

\begin{figure}[h]
\begin{center}
\includegraphics[width = 9 cm]{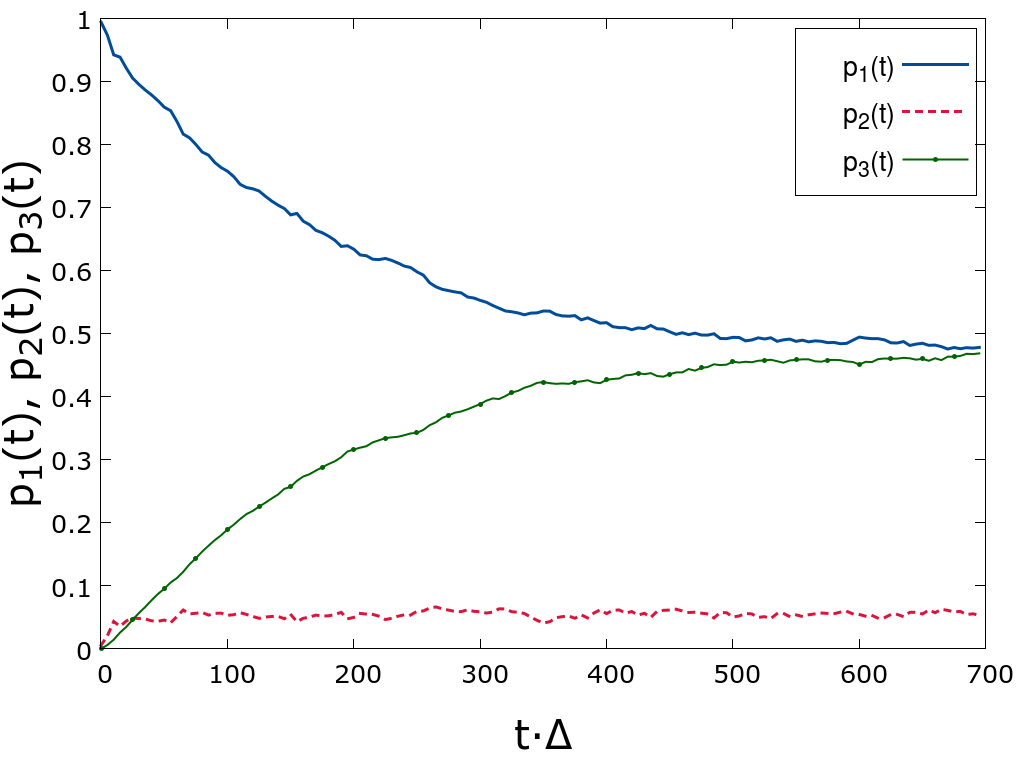}%
\caption{\label{fig7}   Same as in Fig.~\ref{fig5} but for $\epsilon = 17$.}
\end{center}
\end{figure}

As a first result,  we show in Fig. \ref{fig5} - \ref{fig7} the population dynamics of the site occupations $p_j(t)=\text{Tr}\{\ket j \bra j \rho(t)\}$ and $j=\{1,2,3\}$ with $\rho(t_0)=\ket 1\bra 1$ for three different bridge energies $\epsilon = 1$, $\epsilon = 3$ and $\epsilon = 17$ (in units of $\Delta$). For moderate bridge energies one observes coherent transfer of populations towards thermal equilibrium for the chosen coupling parameter $K = 0.24$ and inverse temperature $\beta = 5$, while incoherent (monotonous) decay appears for high-lying bridges. The TCBD captures these qualitatively different dynamical regimes accurately and in domains of parameters space which are notoriously challenging, namely, stronger coupling and very low temperatures.
\subsection*{Rate description}

For the remainder we will demonstrate how the TCBD approach can be used to extract transfer rates in case the population dynamics appears to be incoherent as in Fig.~\ref{fig7}. These rates are of particular relevance for charge and energy transfer in molecular complexes or quantum dot structures, where the three-state system is the simplest model to exhibit sequential hopping from site to site as well as non-local tunneling between donor $|1\rangle$ and acceptor $|3\rangle$. According to Fig.~\ref{fig5a}, one has two different transfer channels, a sequential channel with transfer rate $\Gamma_{\ket 1\rightarrow \ket 2}=\Gamma_{\ket 3\rightarrow \ket 2}=\Gamma_{DB}$ and a super-exchange channel $\Gamma_{\ket 1\rightarrow \ket 3}=\Gamma_{\ket 3\rightarrow \ket 1}=\Gamma_{SQM}$. The respective dominance of these transfer mechanisms allows to classify the reduced system evolution as predominantly classical (hopping) or quantum (tunneling). In the classical limit of high temperatures $\hbar \beta\omega_c \ll 1$ and low bridge state energies $\epsilon$, the former channel is expected to prevail, while for lower temperatures and higher energy barriers, quantum non-locality in both the system and the reservoir degrees of freedom become increasingly important.

Assuming that these two rates define the only relevant time scales leads to a simple population dynamics of the form
\begin{equation}
\begin{pmatrix}
	\dot{p}_1 \\
	\dot{p}_2 \\
	\dot{p}_3
\end{pmatrix} =  \begin{pmatrix}
   -\Gamma_{DB}-\Gamma_{SQM} & \Gamma_{DB} e^{\hbar \beta \epsilon} & \Gamma_{SQM} \\
	\Gamma_{DB} & -2 \Gamma_{DB} e^{\hbar \beta \epsilon} & \Gamma_{DB} \\
	\Gamma_{SQM} & \Gamma_{DB} e^{\hbar \beta \epsilon} & -\Gamma_{DB}-\Gamma_{SQM}
\end{pmatrix}\begin{pmatrix}
	p_1 \\
	p_2 \\
	p_3
\end{pmatrix}\, .
\label{dba}
\end{equation}
 The eigenvalues of the rate matrix are obtained as $\lambda_1 = 0$, $\lambda_2 = -\Gamma_{DB}-2e^{\hbar \beta\epsilon}\Gamma_{DB}$ and $\lambda_3=-\Gamma_{DB}-2\Gamma_{SQM}$ with eigenvectors
\begin{equation}
\vec{v}_1 = \begin{pmatrix}
1 \\ e^{-\hbar\beta\epsilon} \\ 1
\end{pmatrix} \:\:\:\: \vec{v}_2 = \begin{pmatrix}
1 \\ -2 \\ 1
\end{pmatrix} \:\:\:\: \vec{v}_3 = \begin{pmatrix}
-1 \\ 0 \\ 1
\end{pmatrix}
\end{equation}
so that the general solution to (\ref{dba} ) is given by
\begin{equation}
\vec{p}(t) = c_1\vec{v}_1 + c_2 e^{(-\Gamma_{DB}-2e^{\hbar \beta\epsilon}\Gamma_{DB})t}\vec{v}_2+c_3 e^{(-\Gamma_{DB}-2\Gamma_{SQM})t}\vec{v}_3\, .
\label{eigsol}
\end{equation}
Note that due to symmetries and detailed balance one has
\begin{equation}
\Gamma_{DB}(\epsilon) = \Gamma_{BD}(-\epsilon) = \frac{p_B^{\infty}}{p_A^{\infty}}\Gamma_{BD}(\epsilon) = e^{-\hbar\beta\epsilon} \Gamma_{BD}(\epsilon)
\end{equation}
with Boltzmann distributed equilibrium occupation probabilities $p_A^{\infty}$ and $p_B^{\infty}$. Asymptotically,  the dynamics (\ref{dba}) tends towards $p_1^{\infty} = p_3^{\infty}$, $p_2^{\infty} = p_1^{\infty} e^{-\hbar\beta\epsilon} = p_3^{\infty} e^{-\hbar \beta \epsilon}$.

The expression (\ref{eigsol}) constitutes the basis for a numerical extraction of the transfer rates $\Gamma_{DB}$ and $\Gamma_{SQM}$ from simulation data. For this purpose, one introduces auxiliary functions $a(t)$ and $b(t)$ to cast eq. (\ref{eigsol}) in the compact form
\begin{equation}
\vec{p}(t) = \vec{p}_{\infty}+a(t)\cdot \vec{v}_2 +b(t)\cdot\vec{v}_3
\end{equation}
such that the dynamical behaviour of  $a(t)$ and $b(t)$ expressed by the population differences $\tilde{p}_j(t) = p_j(t)-p_j^{\infty}$ is obtained as
\begin{eqnarray}
a(t)&=&\frac{1}{2}[\tilde{p}_1(t)+\tilde{p}_3(t)] \\
b(t)&=&\frac{1}{2}[\tilde{p}_3(t)-\tilde{p}_1(t)]\, .
\end{eqnarray}
The numerical transfer rates $\Gamma_{DB}$ and $\Gamma_{SQM}$ are now extracted by linear least square fits to the logarithm of the auxiliary functions $a(t)$ and $b(t)$. Choosing two parameter fit functions $a_{fit}(t)=c_1^a e^{c_2^a t}$ and $b_{fit}(t)=c_1^b e^{c_2^b t}$, the transition rates can be computed from the fitting parameters as
\begin{eqnarray}
\Gamma_{SQM} &=& \frac{1}{2} \left( \frac{c_2^a}{1+2 e^{\hbar \beta\epsilon}}-c_2^b\right)\\
\Gamma_{DB} &=& - \frac{c_2^a}{1+2 e^{\hbar \beta \epsilon}}.
\end{eqnarray}

\subsection*{Comparison with NIBA rates}

To compare the benchmark results obtained from the TCBD approach with approximate predictions, we come back to the NIBA discussed already in the previous section. As one of the most powerful perturbative treatments, the NIBA has also been the basis to derive analytic expressions for both sequential as well as super-exchange rates in the the domains $\hbar\beta\omega_c \geq 1$ and $\Delta\ll \omega_c$. This way, one arrives at
the sequential forward rate
\begin{equation}
\Gamma_{DB, GR}(\epsilon) = \left(\frac{\Delta}{2}\right)^2\int_{-\infty}^{\infty}dt \exp \left[-i\epsilon t - Q (t)\right]
\label{DB}
\end{equation}
which can also be obtained from a Fermi's golden rule calculation. Upon expanding the dissipative function $Q(t)$ in eq. (\ref{qfun}) to lowest order in $\frac{1}{\hbar\beta\omega_c}$ and $\frac{\epsilon}{\omega_c}$, a simple analytical expression is gained, i.e.,
\begin{equation}
\Gamma_{DB, GR}(\epsilon) = \frac{\Delta^2_{eff}}{4\omega_c}\left( \frac{\hbar\beta\omega_c}{2\pi}\right)^{1-2K} \frac{|\Gamma(K+i\hbar\beta\epsilon/2\pi)|^2}{\Gamma(2K)}e^{\frac{1}{2}\hbar\beta\epsilon}\, ,
\end{equation}
valid for coupling strength $K<1$ and with the effective tunneling matrix element eq. (\ref{tuneff}). Going beyond the second order perturbative treatment to include also fourth order terms in $\Delta$, then leads to an approximate expression for the super-exchange rate
\begin{equation}
\Gamma_{SQM, GR}(\epsilon) \approx \frac{\left(\frac{\Delta}{2}\right)^4}{\epsilon^2}\int_{-\infty}^{\infty}d\tau \exp \left[-4  Q (\tau)\right] \, .
\label{SQM}
\end{equation}

While within the classical, high temperature domain $\hbar\beta\epsilon < 1$ thermally activated processes should be dominant (\ref{DB}) and manifest themselves in the typical Arrhenius behavior $\sim e^{-\beta\epsilon}$, an increase in the bridge energy $\epsilon$ and lower temperature $\beta$ should lead to decay rates according to (\ref{SQM}).
\begin{figure}[h]
\begin{center}
\includegraphics[width = 9 cm]{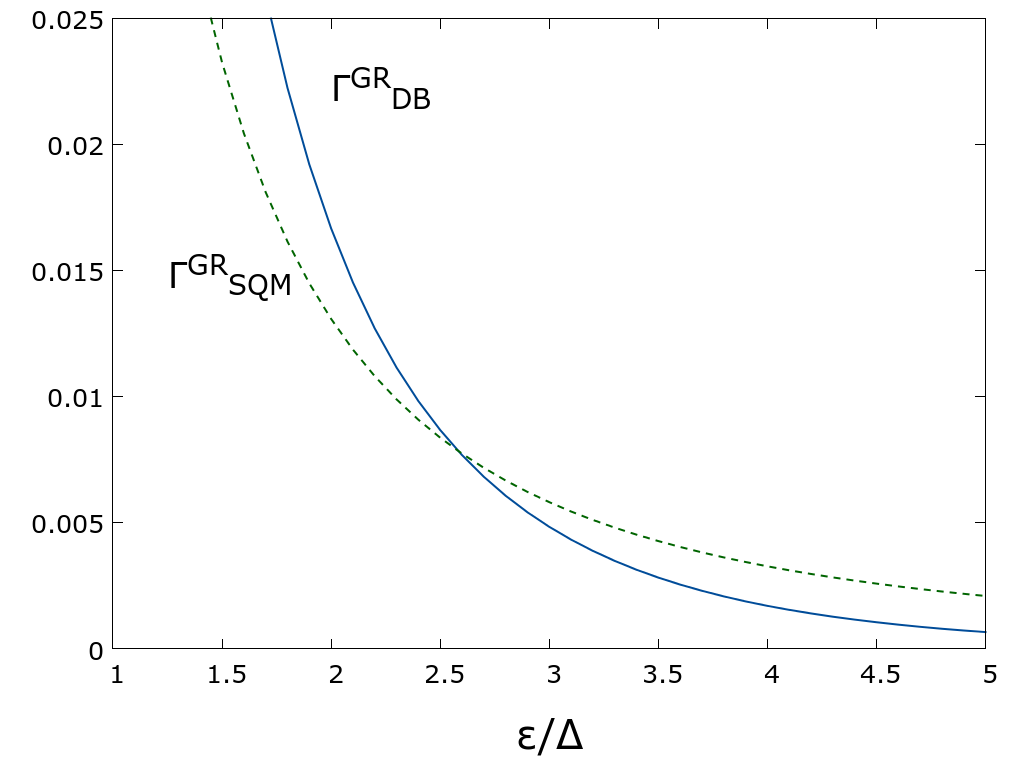}%
\caption{\label{dbsqmcr} Perturbative rates $\Gamma_{DB}^{GR}(\epsilon)$ and $\Gamma_{SQM}^{GR}(\epsilon)$. The changeover from the regime where sequential hopping dominates to the regime of super-exchange occurs around $\epsilon \approx 2.5$; other parameters are $K = 0.24$, and $\beta = 0.7$ (in units of $\Delta$).}
\end{center}
\end{figure}
This is indeed seen in  Fig. \ref{dbsqmcr}, where we depict the rates eq. (\ref{DB}) and (\ref{SQM}) for varying bridge energies.
The regime of sequential transfer crosses over to a regime where quantum tunneling dominates for sufficiently large $\epsilon$, a behaviour that can now be compared to exact numerical results from the TCBD based on  the two rate model of eq. (\ref{dba}).

\begin{figure}[h]
\begin{center}
\includegraphics[width = 9 cm]{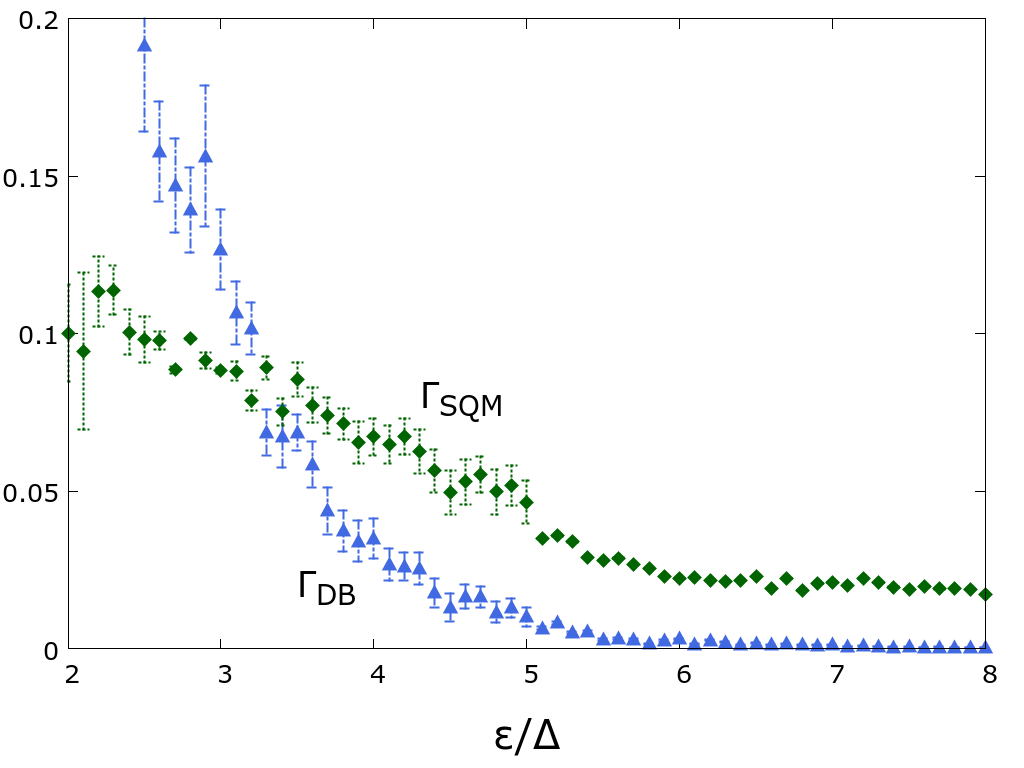}%
\caption{\label{fig10} Rate for sequential hopping $\Gamma_{DB}(\epsilon)$ and super-exchange rate $\Gamma_{SQM}(\epsilon)$ as extracted from TCBD simulations for parameters as in Fig.~\ref{dbsqmcr}. Error bars from the curve fitting procedure are indicated by vertical lines.}
\end{center}
\end{figure}

A comparison between the theoretically predicted classical-quantum crossover in Fig. \ref{dbsqmcr} and the numerical results in Fig. \ref{fig10} reveals at least good qualitative agreement while quantitatively the NIBA rates are not reliable.

\begin{figure}[h]
\begin{center}
\includegraphics[width = 9 cm]{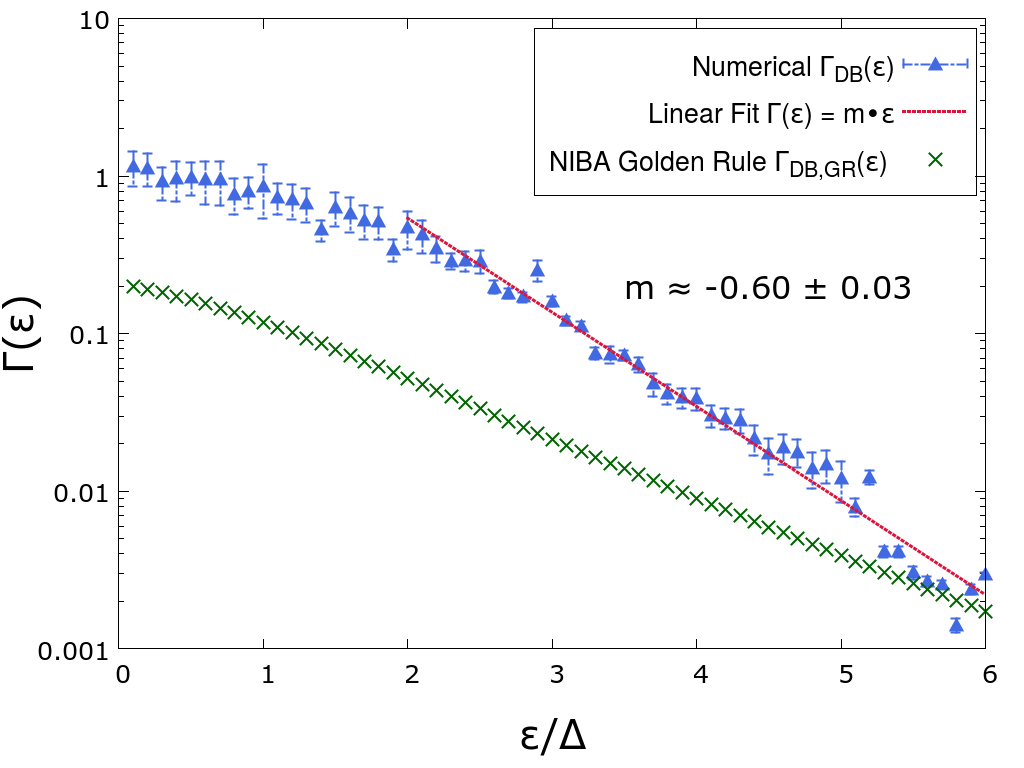}%
\caption{\label{fig11}Sequential hopping rate $\Gamma_{DB}(\epsilon)$ as extracted from TCBD simulations for parameters as in Fig.~\ref{dbsqmcr}. The linear fit $e^{m\cdot\epsilon}$ with slope $m\approx -0.6$ demonstrates good agreement with a thermally activated process with $-\beta\equiv m_{\beta}= -0.7$.}
\end{center}
\end{figure}

In Fig. \ref{fig11} we compare more specifically numerical results for  $\Gamma_{DB}(\epsilon)$ with the NIBA golden rule predictions. The numerical data are in good agreement with an expected Arrhenius behavior, i.e.\ $\Gamma_{DB}(\epsilon)\sim e^{-\beta\epsilon}$, but exceed the NIBA rates quite substantially.

\begin{figure}[h]
\begin{center}
\includegraphics[width = 9 cm]{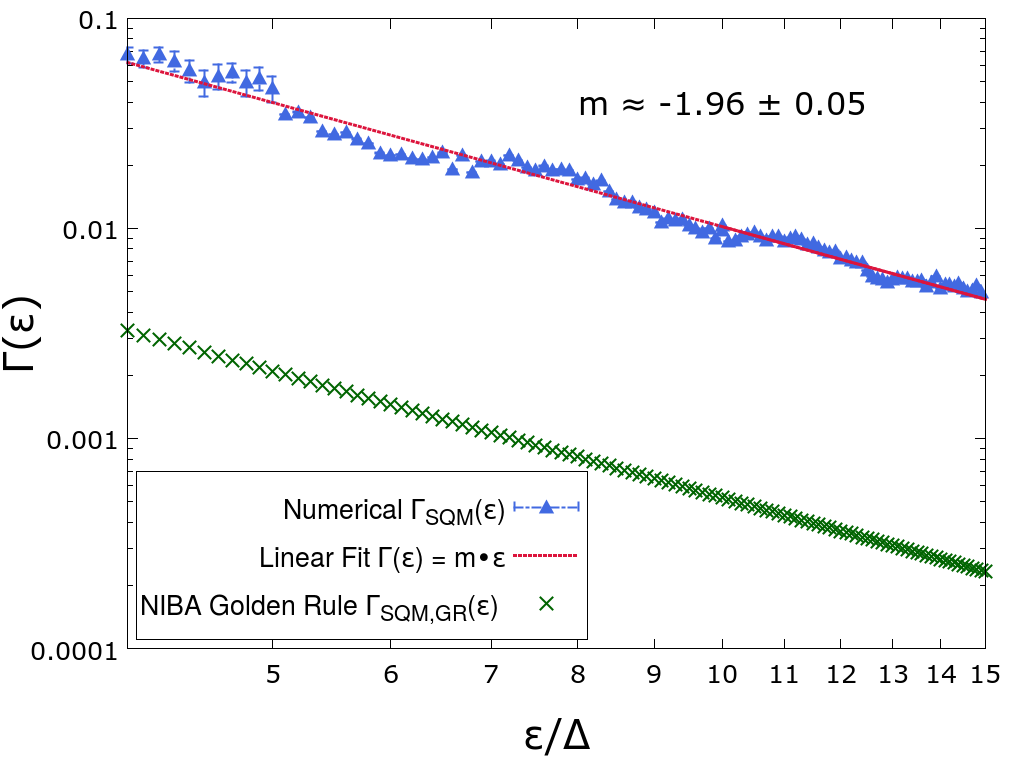}%
\caption{\label{fig12}Super-exchange rate $\Gamma_{SQM}(\epsilon)$ as extracted from TCBD simulation data for parameters as in Fig.~\ref{dbsqmcr}. An algebraic dependence $\frac{1}{\epsilon^{|m|}}$ with $|m| \approx 1.96$ is in agreement with a super-exchange mechanism for which the expected asymptotic prediction is $|m| = 2$.}
\end{center}
\end{figure}

For increasing barrier heights, we expect the super-exchange rate to control the population decay if $\beta\epsilon>1$. The perturbative treatment leads to a characteristic algebraic dependence $\Gamma_{SQM}\sim \frac{1}{\epsilon^{|m|}}$ with $|m|=2$, in contrast to the exponential one for sequential hopping. The results in Fig. \ref{fig12} for moderate to large bridge energies verify the expected power law dependence with an exponent $|m| \approx 1.96$, close to the perturbative expectation.

%\begin{figure}[h]
%\begin{center}
%\includegraphics[width = 9 cm]{highepsdbsqm.png}%
%\caption{\label{fig13}Sequential hopping $\Gamma_{DB}(\epsilon)$ and super-exchange rate $\Gamma_{SQM}(\epsilon)$ as extracted from %simulation data for $K\approx 0.24$, $\Delta = 1$ and $\beta = 0.7$ for high lying intermediate states.}
%\end{center}
%\end{figure}

%\begin{figure}[h]
%\begin{center}
%\includegraphics[width = 9 cm]{highepssqmll.png}%
%\caption{\label{fig14}Super-exchange rate $\Gamma_{SQM}(\epsilon)$ as extracted from TCBD simulation data for very high lying intermediate states (parameters are as in Fig.~\ref{dbsqmcr}). The linear fit $\frac{1}{\epsilon^m}$ with $m \approx -1.9$ is in agreement with the expected asymptotic value $m=-2$.}
%\end{center}
%\end{figure}

%----------------------------------------------------------------------------------------------
%\newpage
\subsection*{Comparison with Master Equation Results}

In case of weak system-reservoir couplings, the master equations build standard perturbative approaches to open system dynamics \cite{breue02}. The corresponding time evolution equations for the reduced density matrix are formulated in terms of the eigenstates of the bare systems with the dissipator inducing transitions between these states. While these methods clearly fail for stronger couplings, it is nevertheless instructive to analyze their deficiencies in the context of rate dynamics.

Here, we use a standard master equation (ME) for which the dynamics of the diagonal elements of the reduced density matrix decouples from that of the off-diagonal elements. In the eigenstate representation of the Hamiltonian eq. (\ref{3ham}), $H_S\ket n = E_n \ket n, \:\:\: n = 1,2,3$, one then has
\begin{equation}
\dot p_n (t)  = \sum_{m = 1}^3 [W_{nm}p_m(t)-W_{mn}p_n(t)]
\label{pma}
\end{equation}
where $p_n\equiv \rho_{nn}$. The transition rates are obtained as
\begin{equation}
W_{mn} = \frac{1}{\hbar^2}\bra m S_z \ket n ^2 \, D(E_m-E_n)
\label{trama}
\end{equation}
with $D(E) = 2 M J(E/\hbar) \bar n (E)$ and the thermal occupation $\bar n (E) = 1/[\exp(\beta E)-1]$. The stationary solution of (\ref{pma}) reproduces the Gibbs state distribution $p_n = Z^{-1}e^{-\beta E_n}$ with partition function $Z$. The dynamics of the off-diagonal elements can simply be solved
\begin{equation}
\rho_{nm}(t)=\rho_{nm}(0)\, {\rm e}^{i(E_n-E_m) t/\hbar}\, {\rm e}^{-\Gamma_{mn}t}
\end{equation}
with decay rates
\begin{eqnarray}
\Gamma_{mn} &=& \frac{1}{\hbar^2} \sum_{r = 1}^3 \frac{1}{2}\left[ \bra m S_z \ket r ^2 D(E_r - E_m) + \bra n S_z \ket r ^2 D(E_r - E_n)\right] \nonumber\\
&&- \frac{1}{\hbar^2} \bra m S_z \ket m \bra n S_z \ket n D(0) .
\label{oddec}
\end{eqnarray}
Now, the population dynamics $P_\mu(t), \mu=1, 2, 3$ in the site representation is obtained from this time evolution by a simple unitary transformation yielding
 \begin{equation}
P_\mu(t) = \sum_{n} c_{\mu n}c^{\ast}_{n \mu} p_{n} + \sum_{n,m} c_{\mu n}c^{\ast}_{m \mu}\rho_{n m} (0) e^{-\frac{i}{\hbar}(E_{n} - E_{m}) t} e^{-\Gamma_{nm}t}\, .
\label{sitra}
\end{equation}
Note that the site populations are also determined by the dynamics of the off-diagonal elements in the eigenstate representation; the respective timescales are entirely determined through eqs. (\ref{trama}), (\ref{oddec}).

\begin{figure}[h]
\begin{center}
\includegraphics[width = 9 cm]{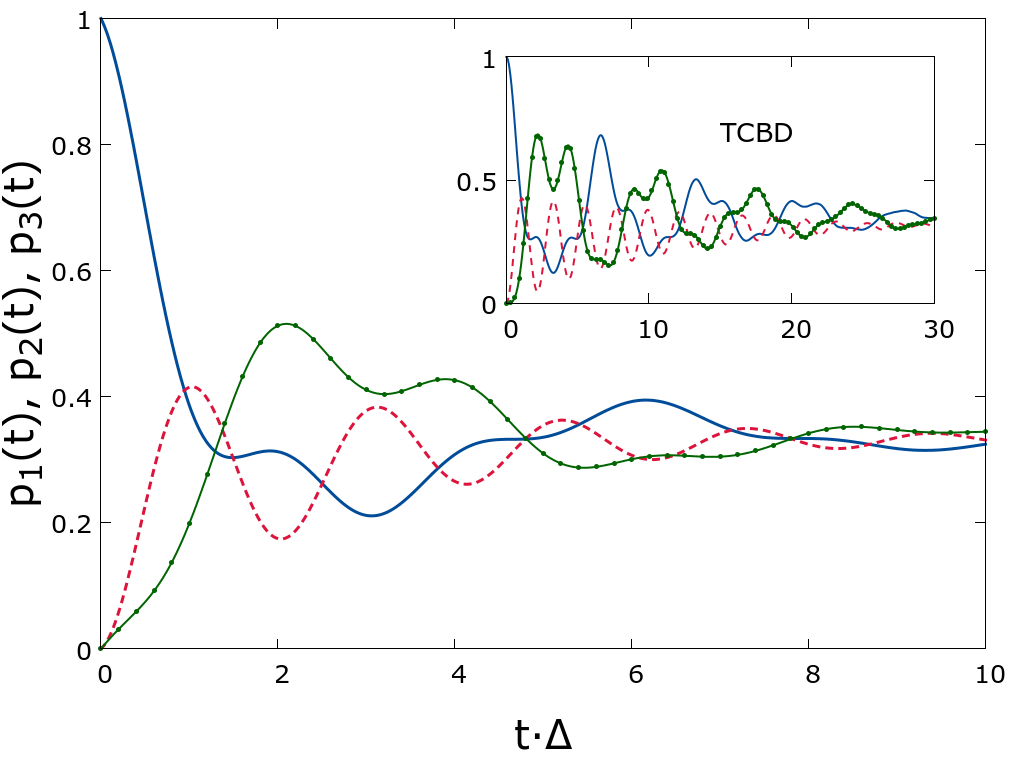}%
\caption{\label{fig15}Population dynamics for DBA complex via ME and TCBD method (inset) for $K = 0.08$, $\beta = 7$, $\omega_c = 50$, and $\epsilon = 1$ (in units of $\Delta$).}
\end{center}
\end{figure}
\begin{figure}[h]
\begin{center}
\includegraphics[width = 9 cm]{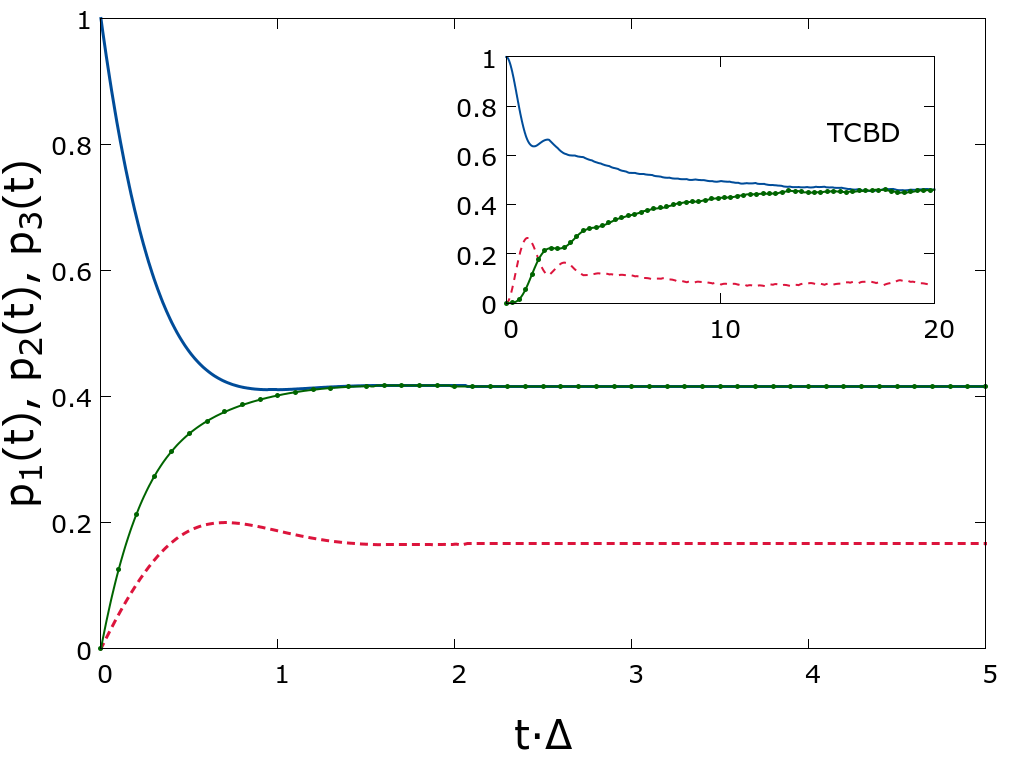}%
\caption{\label{fig16}Same as in Fig.~\ref{fig15} but for $K = 0.24$, $\beta = 0.7$, $\omega_c = 50$, $\epsilon = 2$.}
\end{center}
\end{figure}
%\begin{figure}[h]
%\begin{center}
%\includegraphics[width = 9 cm]{LBsegcomp3.png}%
%\caption{\label{fig17}Population dynamics for DBA complex via Lindblad and TCBD method (inset) for $K \approx 0.08$, $\beta = 5$, %$\omega_c = 50$, $\Delta = 1$ and bias heights $\epsilon = 0$.}
%\end{center}
%\end{figure}

\begin{figure}[h]
\begin{center}
\includegraphics[width = 9 cm]{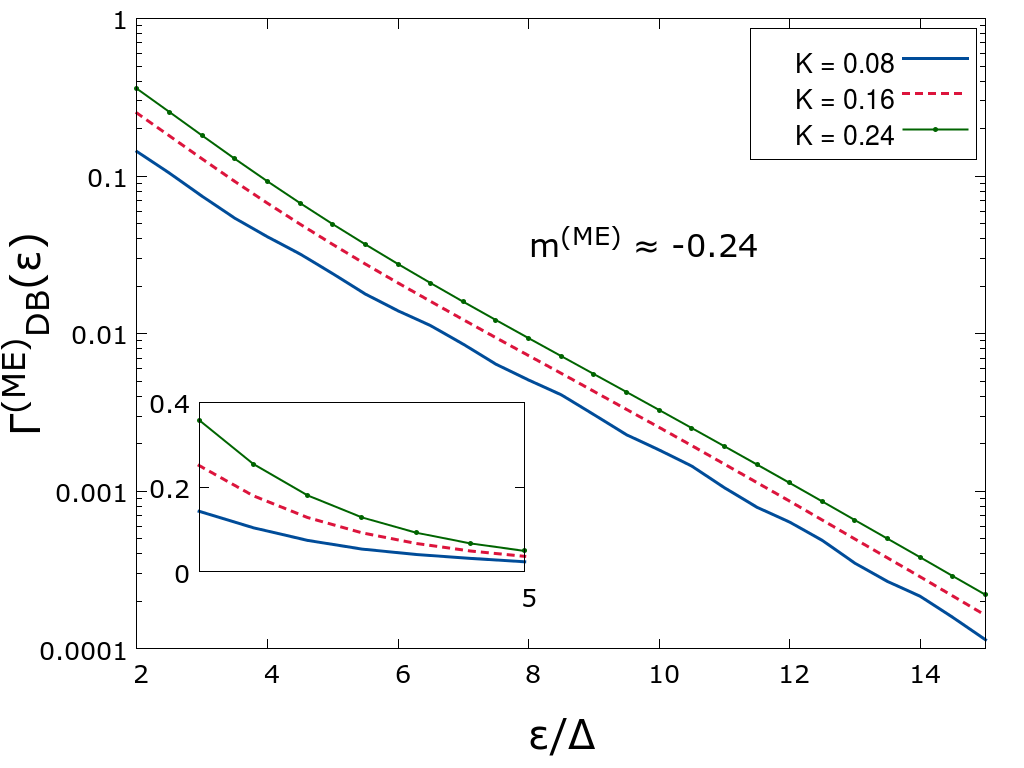}%
\caption{\label{fig18}Sequential transfer rates $\Gamma_{DB}^{(ME)}(\epsilon)$ as derived via the rate model eq. (\ref{dba}) for various damping strengths $K = 0.08,0.16,0.24$ and $\beta = 0.7$ (in units of $\Delta$). A linear fit $e^{m^{(ME)}\cdot\epsilon}$ with $m^{(ME)}\approx -0.24$ reveals strong deviations from the classical expectation for thermal activation $-\beta\equiv m_{\beta}= -0.7$ and the TCBD result $m\approx -0.6$ in Fig. \ref{fig11}.}
\end{center}
\end{figure}

\begin{figure}[h]
\begin{center}
\includegraphics[width = 9 cm]{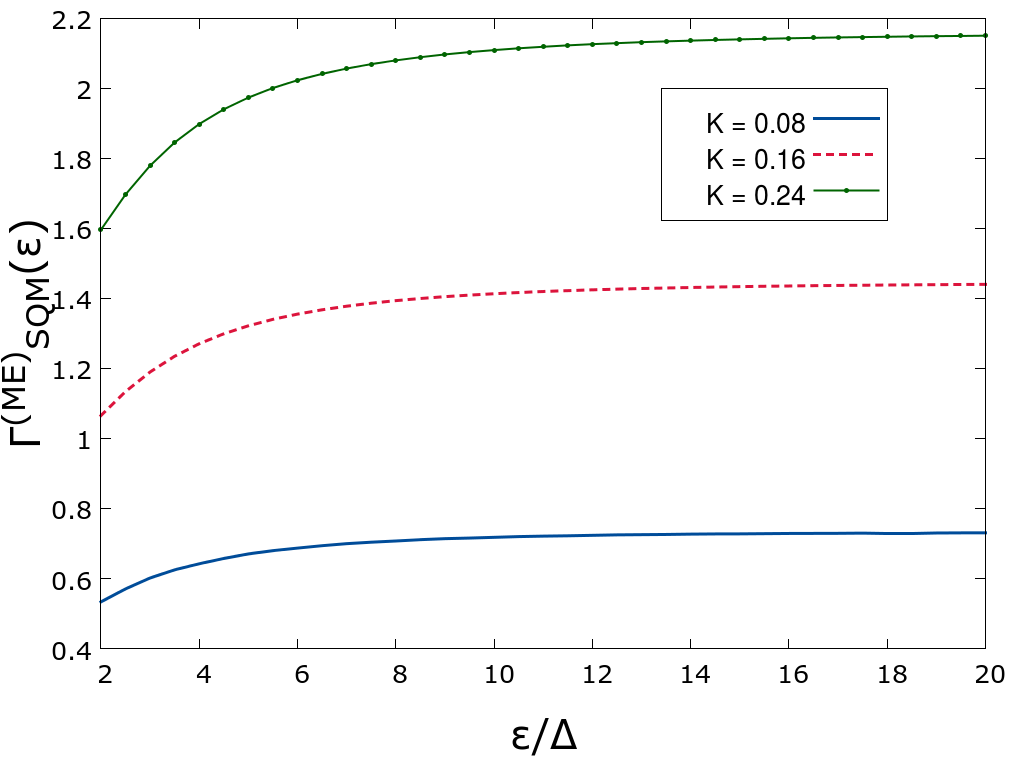}%
\caption{\label{fig19}Same as in Fig.~\ref{fig18} but for the super-exchange rates $\Gamma_{SQM}^{(ME)}(\epsilon)$. These results do not show the expected algebraic dependence on $\epsilon$; see text for details.}
\end{center}
\end{figure}

\begin{figure}[h]
\begin{center}
\includegraphics[width = 9 cm]{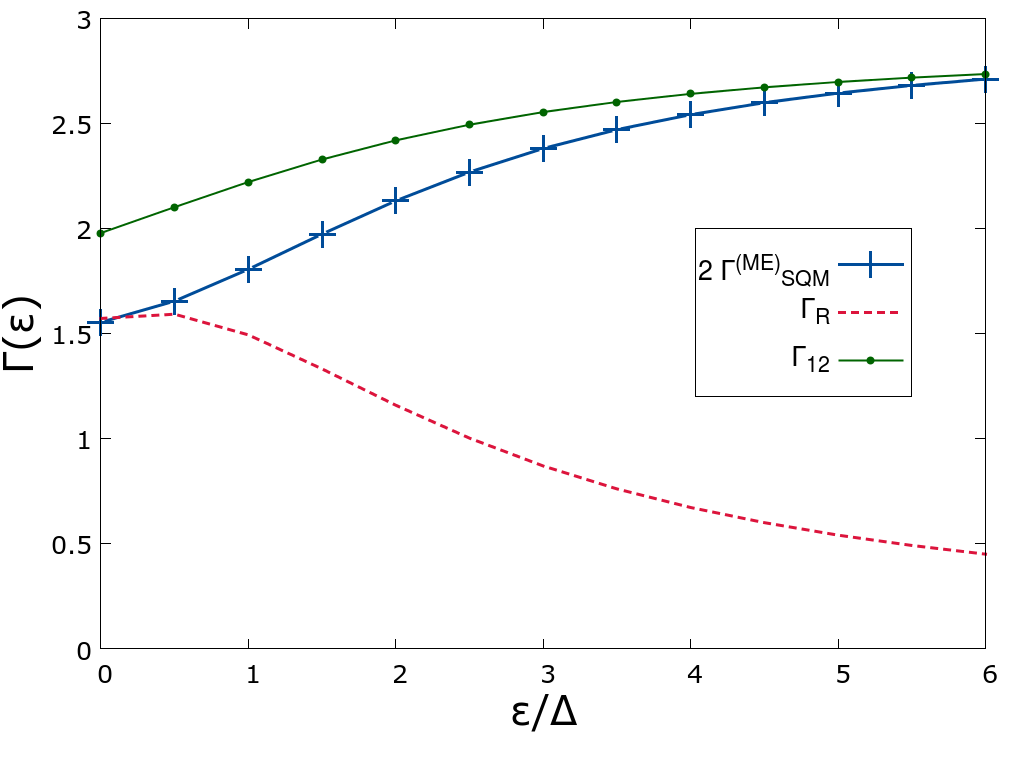}%
\caption{\label{fig20}Decoherence rates according to the ME approach (\ref{pma}) in the energy basis compared to the numerically extracted super-exchange rate $\Gamma_{SQM}^{(ME)}$ in the site representation (\ref{sitra}). While the rate $\Gamma_R$ for energy relaxation
decreases with increasing bridge height, the rate for the decay of coherences in the energy basis (\ref{oddec}) tends to dominate and determines $\Gamma_{SQM}^{(ME)}$; see text for details.}
\end{center}
\end{figure}

We start in Fig. \ref{fig15} with the population dynamics in the coherent regime (weak coupling). While the master results capture the transient oscillatory pattern qualitatively correctly, quantitatively deviations are clearly apparent. The time scales for relaxation towards thermal equilibrium are quite different with the master results approaching a steady state on a much faster time scale than the TCBD data. This may be attributed to the relatively low temperature which is beyond the validity of the Born-Markov approximation on which the master equation is based. The steady state values for the populations are basically identical though. Discrepancies in the relaxation dynamics substantially increase for stronger system bath couplings, see Fig. \ref{fig16}, when the dynamics tends to become a simple decay in time.
%-----------------------------------------------
%-----------------------------------------------
%-----------------------------------------------
Based on the rate extraction method presented before, we can now also extract respective rates from the master equation dynamics providing us with corresponding rates $\Gamma_{DB}^{(ME)}(\epsilon)$ and super-exchange tunneling $\Gamma_{SQM}^{(ME)}(\epsilon)$. The sequential transfer rate in Fig. \ref{fig18} reveals indeed an exponential decay $\sim e^{-|m^{(ME)}|\epsilon}$, however,
with substantial deviations $m^{(ME)}$ from the thermal value $m_\beta$. The situation for the super-exchange rates is even worse as the extracted rates lack a physical interpretation, see Fig. \ref{fig19}: The numerical $\Gamma_{SQM}^{(ME)}(\epsilon)$ saturate for growing bridge energies to eventually become independent of $\epsilon$ at all, in contradiction to an algebraic decay.
%-----------------------------------------------
%-----------------------------------------------
%-----------------------------------------------
%-----------------------------------------------
This result is not so astonishing since, as already mentioned above, the parameter domain where incoherent population dynamics exhibits substantial quantum effects lies at the very edge or even beyond the range of validity of master equations. What is interesting nevertheless, is the fact that the explicit time scales appearing in eq. (\ref{sitra}) can be directly related to the extracted super-exchange rate $\Gamma_{SQM}^{(ME)}(\epsilon)$. This is shown in Fig.~\ref{fig20}. It turns out that indeed non-local processes in the site representation (super-exchange) correspond to the decay rate of off-diagonal elements of the density matrix in the energy representation implying that already for moderate bridge energies $2\Gamma_{SQM}^{(ME)}\to \Gamma_{12}$.

%\newpage
\section{Conclusion}

In this work we combined a stochastic description of open quantum dynamics with projection operator techniques to improve convergence properties. While the general strategy has been outlined recently by one of us \cite{stock16}, we here restricted ourselves to a broad class of systems for which the dephasing time is finite. For the dynamics of the coherences of the reduced density one then keeps track of bath induced retardation effects only within a time window $\tau_m$ which must be tuned until convergence is achieved. While for $\tau_m\to t_{\rm final}$ with $t_{\rm final}$ being the full propagation time one recovers the full stochastic description, the new TCBD  is superior if $\tau_m$ is sufficiently shorter than $t_{\rm final}$. In parameter space this applies to the non-perturbative regime beyond weak coupling, a domain which is notoriously difficult to tackle, particularly at low temperatures.

The new TCBD scheme captures systems with continuous degree of freedom for arbitrary coupling strengths and to those with discrete Hilbert space up to moderate couplings. It allows to approach also the long time domain, where equilibration sets in. For two- and three-level systems we have shown explicitly that the new scheme covers both coherent as well as incoherent dynamics. In this latter regime the method is sufficiently accurate to extract quantum relaxation rates in the long time limit. It now allows for a variety of applications, for example, the dynamics of nonlinear quantum oscillators also in presence of external time dependent driving, the heat transfer through spin chains, or the quantum dynamics under optimal control protocols.

\begin{acknowledgments}
This work was supported by the federal state of Baden-Wuerttemberg through a doctoral scholarship under the postgraduate scholarships act (LGFG) and by the DFG through the SFB/TRR 21.

\end{acknowledgments}

%\nocite{egma93,vochmi89,egma94,egma942,toma94,muehank05,kast13,muehl05}

%%%%%%%%%%%%%%%%%%%%%%%%%%%%%%%%%%%%%%%%%%%%%%%%%%%%%%%%%%%%%%%%%%%%%%%%%%%%%%%%%%%%%

\end{document}